\documentclass[aps,floatfix,superscriptaddress,amsmath,amssymb,noshowpacs,twocolumn]{revtex4}
\usepackage{graphics,lscape,graphicx,epsfig,amsmath,amssymb,epstopdf,wasysym,comment}
\usepackage[T1]{fontenc}
\usepackage{frcursive}
\usepackage{xcolor}
\newenvironment{frcseries}{\fontfamily{frc}\selectfont}{}

\usepackage{subfigure}
\allowdisplaybreaks

\newcommand{\be}{\begin{equation}}
\newcommand{\ee}{\end{equation}}
\newcommand{\bes}{\begin{equation*}}
\newcommand{\ees}{\end{equation*}}
\newcommand{\bea}{\begin{eqnarray}}
\newcommand{\eea}{\end{eqnarray}}
\newcommand{\ba}{\begin{eqnarray*}}
\newcommand{\ea}{\end{eqnarray*}}
\newcommand{\dagga}{{\phantom{\dagger}}}

\newcommand{\bq}{\mathbf{q}}

\newcommand{\bk}{\mathbf{k}}

\newcommand{\dis}{\displaystyle}

\newcommand{\up}{\uparrow}
\newcommand{\down}{\downarrow}
\newcommand{\fract}[2]{\frac{\dis #1}{\dis #2}}

\newcommand{\eqn}[1]{(\ref{#1})}
\newcommand{\ket}[1]{\mid\! #1\rangle}
\newcommand{\bra}[1]{\langle #1\!\mid}

\newcommand{\ep}{{\epsilon}}

\newcommand{\bw}{\begin{widetext}}
\newcommand{\ew}{\end{widetext}}

\newcommand{\nph}{n_\text{e-h}}
\newenvironment{eqs}%
{\begin{equation} \begin{aligned}}%
{\end{aligned} \end{equation} }
\newcommand{\beal}{\begin{eqs}}
\newcommand{\eal}{\end{eqs}}
\newenvironment{eqss}%
{\begin{equation*} \begin{aligned}}%
{\end{aligned} \end{equation*} }
\newcommand{\beals}{\begin{eqss}}
\newcommand{\eals}{\end{eqss}}
\begin{document}

\title{Exciton Mott transition revisited}
\author{Daniele Guerci}
\affiliation{International School for
  Advanced Studies (SISSA), Via Bonomea
  265, I-34136 Trieste, Italy} 
\author{Massimo Capone}
\affiliation{International School for
  Advanced Studies (SISSA), Via Bonomea
  265, I-34136 Trieste, Italy} 
\author{Michele Fabrizio}
\affiliation{International School for
  Advanced Studies (SISSA), Via Bonomea
  265, I-34136 Trieste, Italy} 

\begin{abstract}
The dissociation of excitons into a liquid of holes and electrons in photoexcited semiconductors, despite being one of the first recognized examples of a Mott transition, still defies a complete understanding, especially regarding the nature of the transition, which is found continuous in some cases and discontinuous in others. 
Here we consider an idealized model of photoexcited semiconductors that can be mapped onto a spin-polarised half-filled Hubbard model, whose phase diagram reproduces most of the phenomenology of those systems and uncovers the key role of the exciton binding energy in determining the 
nature of the exciton Mott transition. We find indeed that the transition changes from discontinuous to continuous as the binding energy increases. 
Moreover, we uncover a rather anomalous electron-hole liquid phase next to the transition, which still sustains excitonic excitations despite being 
a degenerate Fermi liquid of heavy mass quasiparticles. 

\end{abstract}

\date{}                                           

\maketitle

\section{Introduction}
The transition between an exciton gas (EG) and an electron-hole liquid (EHL) in photoexcited semiconductors (PES) above the exciton condensation temperature is since long known \cite{Brinkman&Rice-PRB1973,Mott-book,Rice-SSP1978} to realise an almost ideal Mott transition \cite{Mott-1949}, i.e., a metal-insulator transition driven by interaction and not accompanied by any symmetry breaking. 
Nevertheless and despite the great progresses in the theoretical understanding of the Mott transition, several aspects remain puzzling; in the first place the nature of the transition. On one hand, the liquid-gas analogy suggests 
that, as the number of photoexcited electron-hole pairs increases, a gradual crossover between the two phases takes place via the formation, within the EG, of liquid droplets that grow till the system transforms entirely into an EHL, just like in any phase-separation scenario of a first-order transition. However, the concurrent growth of screening might lead to an avalanche effect \cite{Rice-SSP1978} and thus to an abrupt transition into the EHL. This scenario could reveal itself either by the existence of a Mott transition distinct from the gas-liquid one, as Landau and Zeldovich originally proposed for liquid mercury \cite{Landau&Zeldovich}, or through a bistability \cite{Shimano-bistable}. Experimentally, the nature of the transition, which can be studied by photoluminescence or optical absorption, is till now rather controversial. There are, indeed, evidences of two distinct transitions
\cite{Wolfe-PRL1986,Wolfe-PRB1992}, as well as of a bistable behaviour 
\cite{Amo-2007,Stern-PRL2008,Shimano-bistable}, but also of a gradual 
transition \cite{Kappei-PRL2005,Suzuki-PRL2012,Rossbach-PRB2014,Kirsanke-PRB2016}. \\
The exciton physics in semiconductors has been given a new lease 
of life by transition metal dichalcogenides (TMD)
\cite{Wang-NatureNano2012,Duan-ChemSocRev2015,Xiao-Nanoph2016,Yuan-JPCL2017,Wang-RMP2018}, which host strongly bound excitons 
\cite{exciton-binding-dichalcogenides-1,exciton-binding-dichalcogenides-2,exciton-binding-dichalcogenides-3} with appealing potentials \cite{Mueller-npj2018}, e.g., in spin- and valley-tronics 
\cite{Mak-NatNanotec2012,Li-PRL2014,Shimazaki-NatPhys2015,Jin-Science2018}. 
The dynamics of photoexcited electron-hole pairs in monolayer TMD 
has been investigated in many experiments, 
see, e.g., Refs.~\cite{Chernikov-NaturePhotonics2015,Pogna-ACSNano2016,Steinleitner-NanoLett2017,Cunningham-ACSNano2017,Sie-NanoLett2017,Ruppert-NanoLett2017,Jiang-Opt2018,Bataller-NanoLett2019}. 
At low excitation density, there is consensus that both the excitons 
and the electronic gap red shift 
\cite{Pogna-ACSNano2016,Cunningham-ACSNano2017,Sie-NanoLett2017}, signalling 
an important contribution of Coulomb repulsion in the semiconducting state 
of TMD. At higher excitation densities, where the EG-EHL transition is 
expected to occur, the situation is less clear. Reflectance measurements  
in WS$_2$ \cite{Chernikov-NaturePhotonics2015} irradiated by an ultrashort laser pulse show a gradual bleaching of the exciton 
absorption peak and, at lower energies, a loss of reflectance that is  
attributed \cite{Chernikov-NaturePhotonics2015} to an EHL phase with a more than 20\% reduction of the gap. The 
coexistence of both signals indicates phase separation, and thus a continuous 
transformation from the EG to the EHL. On the contrary, time-resolved 
photoluminescence in MoS$_2$ during a long 500~ns pulse photoexcitation  \cite{Bataller-NanoLett2019} reveals, at low pump fluence, the aforementioned red shift and a broadening of the exciton emission peak, which, above a 
threshold fluence, suddenly turns into a much broader 
and five times more intense emission peak, centred 200 meV below. This behaviour is rather suggestive of a discontinuous transition, unlike what observed in WS$_2$ \cite{Chernikov-NaturePhotonics2015}. Surprisingly, 
photoluminescence stops right after the 500~ns pump pulse \cite{Bataller-NanoLett2019}, which is interpreted as the system undergoing a transformation from direct to indirect gap semiconductor, possibly driven by lattice expansion. 
All this suggests that photoexcited TMD may show rather interesting properties, especially because of the important role played by Coulomb 
repulsion.   \\  
  
In view of the revived interest in the physics of excitons, the solution of the basic yet open issues in the exciton Mott transition cannot be further delayed. The scope of the present work is just putting together some pieces of that puzzle. For that purpose, we consider an idealised model of 
PES that can be mapped onto a half-filled repulsive Hubbard model at finite spin polarisation, where the fully polarised state, a trivial insulator, maps onto the unexcited semiconductor, and each spin flip corresponds to adding one electron and one hole in the conduction and valence bands, respectively.  
In turns, the insulator-metal Mott transition reached at large enough Hubbard $U$ upon decreasing spin polarisation translates into the EG-EHL 
transition on increasing the density of photoexcited electron-hole pairs. 
We find that such Mott transition can be either continuous or discontinuous, 
in the sense specified above, depending on the strength of $U$, which translates into the magnitude of the exciton binding energy, much in accordance with experiments. 

The plan of the paper is as follows. In Sec. \ref{model} we introduce a simple modelling of PES, while in Sec. \ref{method} the tools we make use to study the model. The phase diagram that we thus obtain is discussed in Sec. \ref{Mott_PES}. Finally, Sec.~\ref{Conclusions} is devoted to concluding remarks. 
 
\begin{figure}[htb]
\begin{center}
\includegraphics[width=0.45\textwidth]{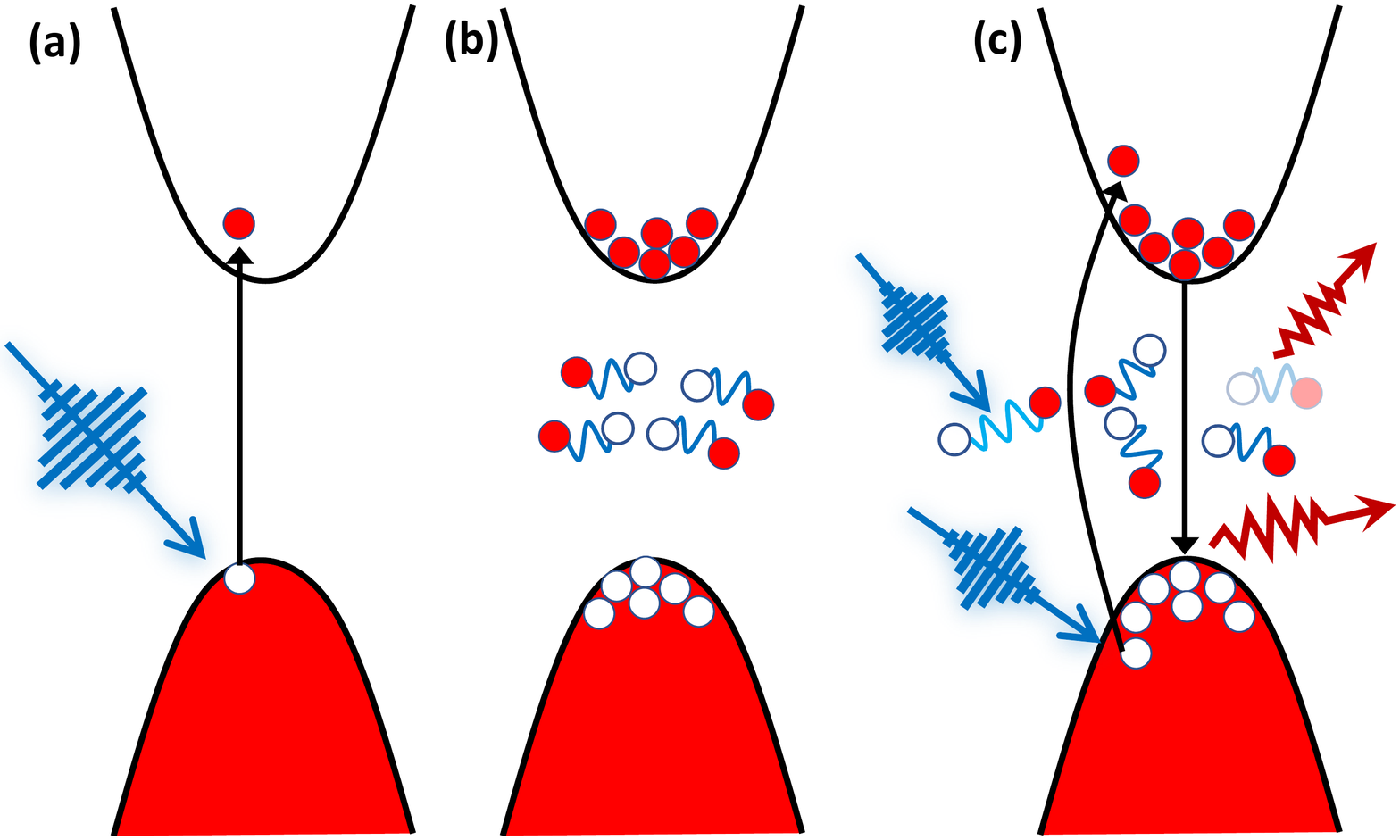}
\vspace{-0.15cm}
\caption{\textbf{Photoexcitation of a semiconductor by a laser pulse}. (a) A semiconductor initially at equilibrium is shot by a laser pulse. (b) A transient quasi-stationary local equilibrium state is established with excitons in the gap, and electron and holes 
in the conduction and valence bands, respectively. (c) Such local equilibrium state can be probed by optical absorption of light, blue beams, which either transfers additional electrons across the gap, above an absorption edge blue shifted by the presence of other electrons and holes, or excites internal states of the exciton. Alternatively, photoluminescence can be used as probe, red beams, which corresponds to radiative recombination either of bound or unbound electron-hole pairs.}
\label{toy_model}
\end{center}
\end{figure}
  
\section{The Model}
\label{model}

In Fig.~\ref{toy_model} we describe schematically a PES in the simple case of
a direct-gap single-valley semiconductor. A laser pulse excites electrons 
across the gap, thus leaving behind holes in the valence band and creating electrons in the conduction one, panel (a). If the electron-hole (e-h) recombination time is long enough, a quasi-stationary local-equilibrium state sets in, panel (b), at finite densities of electrons, $n_\text{e}$, and holes, $n_\text{h}$, which, at low temperature, are equal to the density $\nph$ of photoexcited e-h pairs, i.e., $n_\text{e}=n_\text{h}=n_\text{e-h}$. Some of  them binds together and form excitons, drawn inside the gap 
as e-h pairs connected by strings, while others remain unbound. Such quasi-stationary state can be probed either by absorption or photoluminescence, blue and red light beams in panel (c). Specifically, the system can 
absorb light by creating additional e-h pairs above an absorption edge shifted by the presence of already existing particles and holes, or, at lower energy, through intra-exciton transitions \cite{Kaindl-Nature2003}. 
Photoluminescence is expected to arise by the radiative recombination both of unbound e-h pairs and of excitons. These two processes emit at different frequencies, and thus the corresponding emission intensities 
gives a measure of the relative populations of bound and unbound e-h pairs. 
When the gap is instead indirect, light absorption and emission must be accompanied by emission of phonons to compensate the momentum mismatch. 

The quasi-stationary state in the simple case of Fig.~\ref{toy_model} 
can be described by the Hamiltonian  
\beal
\label{attractive_model}
H &= \sum_{\bk\sigma}\Big(\epsilon_{h\bk}\,
h^\dagger_{\bk\sigma}\,h^\dagga_{\bk\sigma}+
\ep_{e\bk}\,e^\dagger_{\bk\sigma}\,e^\dagga_{\bk\sigma}\Big)\\
&+\sum_{\bq}\,\frac{U(\bq)}{2V}\,\Big(\rho_{h\,\bq}-\rho_{e\,\bq}\Big)
\Big(\rho_{h\,-\bq}-\rho_{e\,-\bq}\Big)\,,
\eal 
at fixed and equal densities of particles and holes, $n_\text{e}=n_\text{h}=n_\text{e-h}$. The operators $h^\dagger_{\bk\sigma}$ and 
$e^\dagger_{\bk\sigma}$ create, respectively, a hole in the valence band, 
with energy cost $\ep_{h\bk}$, and an electron in the conduction one, 
with energy cost $\ep_{e\bk}$, both with momentum $\bk$ and spin $\sigma$.  
$U(\bq)$ is the Coulomb interaction screened by all bands but valence and conduction ones, 
while $\rho_{h\,\bq}$ and $\rho_{e\,\bq}$ the densities at momentum $\bq$ of holes and electrons, which have opposite charges. 
Because of our assumption of quasi-stationarity, we do not include in \eqn{attractive_model} recombination processes, so that 
$n_\text{e}$ and $n_\text{h}$ are separately conserved.  
 
In order to single out the interaction physics, we consider here an idealised modelling, discussed, e.g., in Ref.~\cite{Nozieres1985}, obtained by further simplifying the Hamiltonian \eqn{attractive_model}. 
First, since the e-h Coulomb attraction is primarily a charge effect, we ignore the spin, and thus assume spinless holes and particles. Second, we neglect the effective mass difference between valence and conduction bands, implying $\ep_{h\bk}=\ep_{e\bk}\equiv \ep_\bk$, and, for simplicity, assume the latter as the dispersion relation of a tight-binding model with nearest neighbour hopping, which is also quadratic in the 
small-$\bk$ regime pertinent to low-density. Finally, we replace the long-range Coulomb interaction 
$U(\bq)$ by a short range one, $U$, constant in $\bq$, so that the Hamiltonian \eqn{attractive_model} 
transforms into  
\beal
\label{attractive_model_1}
H &= \sum_{\bk\sigma}\epsilon_{\bk}\,\Big(
h^\dagger_{\bk}\,h^\dagga_{\bk}+
e^\dagger_{\bk}\,e^\dagga_{\bk}\Big)\\
&-U\sum_{i}\,n_{e\,i}\,n_{h\,i}-\mu\sum_{i}\big(n_{e\,i}+n_{h\,i}\big)\,,
\eal
where $n_{h(e)\,i}$ is the local density at site $i$ of holes(electrons), and the chemical potential $\mu$ is such as to fix $\langle n_{h\,i}\rangle=\langle n_{e\,i}\rangle=\nph$, where $\nph\ll 1$ is the density of 
photoexcited e-h pairs. We remark that the model 
\eqn{attractive_model_1} can describe the same physics of \eqn{attractive_model} only if  $U$ is large enough to create bound states below the two-particle continuum, which play the role of the excitons in the original system. 

The simplified Hamiltonian \eqn{attractive_model_1} can be mapped onto a standard repulsive Hubbard model 
\beal
H&= \sum_{\mathbf{k}}\sum_{\sigma}\,\epsilon_{\mathbf{k}}\,
d^\dagger_{\mathbf{k}\sigma}\,d^\dagga_{\mathbf{k}\sigma}\\
&+U\,\sum_i\,n_{i\up}\,n_{i\down}-h\,\sum_i\, \big(n_{i\up}-n_{i\down}\big)
\,,
\label{Ham}
\eal
through the transformation 
\beal
\label{PH_transf}
e^\dagga_{i}&\to d^\dagga_{i\downarrow}\,,&	
h^\dagga_{i}&\to (-1)^{i}\,d^\dagger_{i\uparrow}\,,
\eal
under which $n_{h\,i} \to 1 - n_{i\up}$, and 
$n_{p\,i} \to  n_{i\down}$, so that 
 \beal
2\nph&= \langle n_{h\,i}\rangle + \langle n_{e\,i}\rangle
\to 1 - \langle n_{i\up}\rangle + \langle n_{i\down}\rangle = 1-m\,,\quad\\
0 &= \langle n_{h\,i}\rangle - \langle n_{e\,i}\rangle \to 
1 -\langle n_{i\up}\rangle - \langle n_{i\down}\rangle = 1-n\,.
\label{mapping-PES}
\eal 
It follows that the model \eqn{Ham} must be studied at half-filling, $n=1$, and, in addition, at fixed magnetisation $m=1-2\nph$, which can be enforced by the Lagrange multiplier $h$ playing the role of a fictitious Zeeman field. 

Therefore, the physics of PES can be captured by the half-filled repulsive Hubbard model at fixed, and large if  $\nph\ll 1$, magnetisation $m$, provided all assumptions above are valid. 
Indeed, the Hamiltonian \eqn{Ham} is expected to display several phases in one to one correspondence to those of PES \cite{Massimo02,Scalapino2007,Lieb89,Rok_Osolin_2015}: a low-temperature canted antiferromagnetic insulator, which translates into a phase of condensed excitons, 
and high temperature paramagnetic Mott insulating and metallic phases, which correspond to the EG and EHL, respectively. 
We are particularly interested in the high-temperature phases and the transition between them, which can be inferred from ground-state calculations of \eqn{Ham} if we force the translational and residual spin-$U(1)$ symmetries, on the proviso that preventing symmetry breaking in a ground-state calculation correctly predicts which phases occur above the critical temperature.

\section{The method}
\label{method}

The Hamiltonian \eqn{Ham} is ideally suited to dynamical mean-field theory (DMFT) \cite{ReviewDMFT}, which has emerged as one of the state-of-the-art methods to study Mott transitions, especially when they are not associated to spatial ordering. However, earlier results on the simple half-filled repulsive Hubbard model in a Zeeman field \cite{GeorgesPRB1994,Metzner01,Massimo02,HewsonPRB2007,Zhu_2017} show some differences in the nature of the transition, despite an overall agreement on the main features of the phase diagram.
\begin{figure}
\begin{center}
\includegraphics[width=0.42\textwidth]{{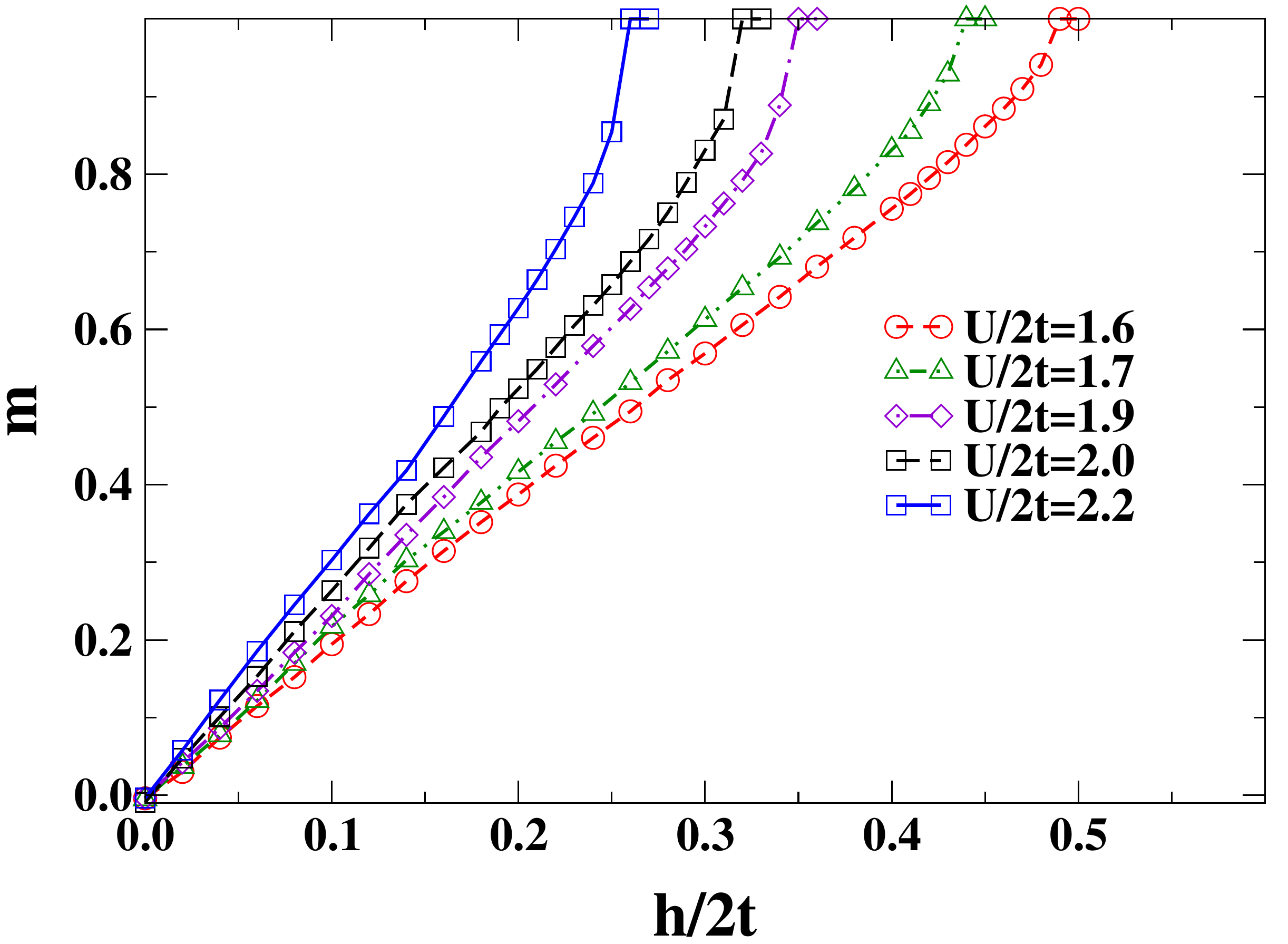}}
\vspace{-0.15cm}
\caption{Magnetisation $m$ vs. $h$ for increasing values of $U/2t$ obtained by DMFT using exact diagonalization at zero temperature with 8 bath's sites, and 
a semicircular density of states of width $4t$. In the large $U$ regime relevant to PES, we always find 
upon increasing $h$ a first order transition from a partially polarised metal to a fully polarised insulator. }
\label{m_vs_h}
\end{center}
\end{figure} 
In particular, in the large $U\gg h$ regime pertinent to PES, the Mott transition is always found first order, but Refs.~\cite{GeorgesPRB1994} and \cite{Massimo02} predict a transition between a partially polarised metal and a partially polarised insulator, while Refs.~\cite{HewsonPRB2007} and \cite{Zhu_2017} report a transition between a partially polarised metal and a fully polarised insulator, and no evidence of a partially polarised insulator. We also performed a zero-temperature DMFT calculation using exact diagonalization as impurity solver, and we could not stabilise a partially polarised insulator, see the magnetisation $m$ vs. $h$ shown in Fig. \ref{m_vs_h}, in agreement with \cite{HewsonPRB2007} and \cite{Zhu_2017}. 
 	
The reason of this discordance can be readily traced back to the iterative scheme employed to solve DMFT, 
which fails to converge when forcing the translational symmetry and the residual $U(1)$ spin-rotational symmetry around the $z$-axis parallel to  
$h$, both of which are instead spontaneously broken in the true canted antiferromagnetic ground state.  \\  
In DMFT, assuming a semicircular density of states (DOS) and forcing 
the aforementioned symmetries, the lattice model \eqn{Ham} is mapped onto an Anderson impurity model (AIM), where the spin-resolved hybridisation function with the bath, $\Gamma_\sigma(\ep)$, is self-consistently determined by the single-particle DOS of the impurity, $\mathcal{A}_\sigma(\ep)$ \cite{ReviewDMFT} -- which in turn must correspond to the local DOS of the lattice model -- through the equation 
\be
\Gamma_\sigma(\ep) = t^2\,\mathcal{A}_\sigma(\ep)\,.\label{DMFT-cond}
\ee 
Imagine we solve iteratively Eq.~\eqn{DMFT-cond} at $h\ll t,U$ 
in the Mott insulating regime, which translates into an AIM whose 
hybridisation function $\Gamma(\ep)= \Gamma_\up(\ep)+\Gamma_\down(\ep)$ has a gap of order $U$ at the chemical potential. This implies that, at any iteration $i$, 
the hybridisation function is obtained by the impurity DOS at the previous iteration, 
$\Gamma^{(i)}_\sigma(\ep)= t^2\,\mathcal{A}^{(i-1)}_\sigma(\ep)$, and thus acquires the same spin polarisation. Since the impurity-bath hybridisation entails an effective antiferromagnetic coupling, at a given iteration the spin of the impurity will be polarised in the opposite direction of that of the bath,  thus opposite to the same impurity at the previous iteration. The reversal of the impurity spin from an iteration to the next one is unavoidable when $\Gamma(\ep)$ is gapped, unless $h$ is large enough to prevail over the effective antiferromagnetic coupling with the bath and thus align the impurity spin parallel to it. If so, at  the next iteration also the bath will be aligned with $h$, and the iterations will converge to 
a trivial solution where both bath and impurity are fully spin polarised along $h$, which is evidently eigenstate of the AIM. In other words,  
the iterative procedure in the Mott insulating phase either does not converge to any solution at all or, for 
$h$ above a threshold, it converges to a trivial solution that represents a fully spin-polarised 
insulator, in accordance with our calculations and Refs. ~\cite{HewsonPRB2007} and \cite{Zhu_2017}, or, 
in the language of PES, to the equilibrium state without photoexcited e-h pairs. \\
The lack of convergence at small $h$, which prevents the stabilisation of a partially polarised insulator and simply signals the tendency to form an antiferromagnetic state, can be formally avoided by choosing a convergence criterium only within same-parity iterations, $i$, $i+2$, \ldots , and, at the end, assuming 
as impurity DOS, $\mathcal{A}_\sigma(\ep)$, the average of those at even and odd iterations~\cite{Massimo02}. This choice is equivalent to assume a mixed state, despite the temperature is zero, where the partial spin polarisation of the Mott insulator results from a statistical ensemble of pure states, with the impurity spin polarised in opposite directions.  
Since the impurity spin polarisation maps in DMFT into the lattice magnetisation, $m$, which, in turn, translates into $1-2\nph$ in the language of PES, see Eq.~\eqn{mapping-PES}, such statistical ensemble of states with $1-2\nph=m\sim +1$ and $1-2\nph=m\sim -1$  
actually describes a rather strange, and not very physical, phase-separated exciton Mott insulator, where the bound e-h pairs are circumscribed within a finite portion of the system, $\nph\sim 1$, while they are almost absent in the rest, $\nph\sim 0$.   
This result would not change when solving, still iteratively, the DMFT 
self-consistency equation \eqn{DMFT-cond} at small but finite temperatures~\cite{GeorgesPRB1994}.
 
\subsection{The variational ghost-Gutzwiller wavefunction}
 \label{g-GW_vt}

We emphasise once more that the issue here is the iterative implementation of 
DMFT when forcing symmetries that are instead spontaneously broken in the 
true ground state. For instance, we would not expect to find the same 
unsatisfactory results in a direct constrained optimisation of the DMFT functional \cite{Kotliar1999}, which however has never been implemented in practice. 
A less rigorous approach, but equally devoid of the problems outlined before,   
would be the minimisation of the expectation value of the Hamiltonian on a constrained variational wavefunction, forced to be invariant under the 
aforementioned symmetries. In what follows we shall adopt just this variational approach. 

The main difficulty in this context is finding a variational wavefunction that can faithfully describe a Mott insulator. We shall use here a recent  extension~\cite{Lanata-ghost-2017} of the Gutzwiller wavefunction, which has been named \textit{ghost} Gutzwiller wavefunction (g-GW) and reads  
\be
\ket{\Psi} = \prod_i\,\mathcal{P}_i\ket{\Psi_*}\,,
\label{G-wf}
\ee
where $|\Psi_*\rangle$ is a Slater determinant, ground state of a tight-binding variational Hamiltonian $H_*$ for $N\geq 1$ spinful orbitals, while 
\be
\mathcal{P}_i = \sum_{\Gamma\gamma}\,\lambda_{\Gamma\gamma}(i)\,
\ket{\Gamma_i}\bra{\gamma_i}\,,
\ee  
is a linear map at site $i$, parametrised by the 
variational parameters $\lambda_{\Gamma\gamma}(i)$, from the local $N$-orbital Hilbert space, 
spanned by the states $\ket{\gamma_i}$, to the physical single orbital 
local space, spanned instead by the states $\ket{\Gamma_i}$. 
In what follows, we shall denote as $c^\dagga_{ia\sigma}$ 
and $c^\dagger_{ia\sigma}$, $a=1,\dots,N$, the annihilation and creation 
operators, respectively, of the $N$ orbitals per each site $i$ in the enlarged 
Hilbert space, while as $d^\dagga_{i\sigma}$ 
and $d^\dagger_{i\sigma}$ the operators of the physical orbital.  
We note that 
for $N=1$ the g-GW in \eqn{G-wf} reduces to the conventional Gutzwiller 
wavefunction~\cite{Gutzwiller-1,Gutzwiller-2}. 

The trick of adding subsidiary degrees of freedom to improve the accuracy of a variational wavefunction has a long history that goes back to the shadow wavefunctions for $^4$He \cite{Shadow-WF}, and is still very alive, as testified by the great interest in matrix-product states and tensor networks \cite{MPS-1,MPS-2}, or, more recently, in neural-network quantum states \cite{Carleo-Science2017}.\\
The wavefunction \eqn{G-wf} is simple and bears close similarity to DMFT. 
Indeed, as noted in~\cite{NicolaPRX}, the variational parameters $\lambda_{\Gamma\gamma}(i)$ can be associated to the components $\psi_i(\Gamma,\bar{\gamma})$, where 
$|\bar\gamma\rangle$ is the particle-hole transform of $|\gamma\rangle$, of a wavefunction 
$\ket{\psi_i} = \sum_{\Gamma\bar\gamma}\, \psi_i(\Gamma,\bar{\gamma})
\,|\Gamma\rangle\!\times\!|\bar\gamma\rangle$ that describes an impurity 
at site $i$ coupled to a bath of $N$ levels. The analogy with DMFT is thus self-evident. Hereafter, in order to enforce translationally symmetry and the $U(1)$ symmetry under spin rotation around the $z$-axis, we shall assume that 
$\ket{\Psi_*}$ is translationally invariant, 
$\ket{\psi_i} = \ket{\psi}$, $\forall\,i$, and that both wavefunctions are eigenstates of the total $z$-component of the spin.

The expectation values on the g-GW can be computed analytically in 
lattices with infinite coordination~\cite{Lanata-ghost-2017}, 
just like it happens for the conventional Gutzwiller 
wavefunction~\cite{BW&G1998}, although it is common to use the 
same analytical expressions also in lattices with finite coordination,
just replacing the corresponding tight-binding DOS, which goes under the name of Gutzwiller approximation. For that, one needs to impose    
in the impurity model representation~\cite{NicolaPRX}, the obvious normalisation condition  
\be
\label{GW_constraints_1}
\bra{\psi}\psi\rangle=1,
\ee
as well as the additional constraint 
\be
\label{GW_constraints_2} 
\bra{\psi}c^\dagga_{b\sigma}\,c^{\dagger}_{a\sigma}\ket{\psi}=
\bra{\Psi_{*}}c^\dagger_{ia\sigma}\,c^\dagga_{ib\sigma}\ket{\Psi_{*}}
\equiv \Delta_{\sigma,ab}\,, 
\ee
where the fermionic operators on the l.h.s. refer to the $N$ levels of the bath coupled to the impurity. We remark that $\Delta_{\sigma,ab}$ 
is independent of $i$ since $\ket{\Psi_{*}}$ is by assumption translationally invariant, and diagonal in spin 
since $\ket{\psi}$ and $\ket{\Psi_{*}}$ are both spin $U(1)$ invariant.         
 
Another important ingredient to compute expectation values 
is the wave-function renormalization vector $\mathbf{R}_\sigma$ with elements $R_{a\sigma}$, 
obtained by solving the set of equations: 
\beal
\label{R_factors}
\bra{\Psi_{*}}c^\dagger_{ia\sigma}\,\mathcal{P}^\dagger_{i}\,
d^\dagga_{i\sigma}\,\mathcal{P}_{i}\ket{\Psi_{*}}&=\sum_{b}\,
\Delta_{\sigma,ab}\,R_{\sigma,b}\\
&\equiv \Big(\hat\Delta_\sigma\,\mathbf{R}_\sigma\Big)_a
\,,\quad
\eal
whose formal solution reads~\cite{Michele17}
\be
\label{wavefunction_Rf}
\mathbf{R}_\sigma = \hat{S}^{-1}_\sigma\,\mathbf{Q}_\sigma\,,
\ee
where the vector $\mathbf{Q}_\sigma$ has components 
\be
Q_{\sigma,a}=\bra{\psi}c^\dagger_{a\sigma}\,d^\dagga_{\sigma}\ket{\psi}\,,
\ee
and the Hermitian matrix $\hat{S}_\sigma$ is defined through  
\be
\hat{S}^2_\sigma=\hat{\Delta}_{\sigma}\cdot\left(1-\hat{\Delta}_{\sigma}\right).
\ee

We apply the g-GW approach to study the model \eqn{Ham} on a Bethe lattice 
with infinite coordination, which corresponds to a semicircular tight-binding DOS. We choose to work with $N=3$ subsidiary orbitals, which was 
shown \cite{Lanata-ghost-2017} to provide already very accurate ground state properties in comparison with DMFT. 
Moreover, we treat all components $\psi(\Gamma,\bar\gamma)$ of the wavefunction describing the quantum impurity coupled to the bath of $N=3$ levels as free variational parameters, apart from the normalisation and the constraints imposed by spin $U(1)$ and particle-hole symmetries. 
This means that, in contrast with the original work~\cite{Lanata-ghost-2017},  we do not require $\psi(\Gamma,\bar\gamma)$ to be ground state of an auxiliary Anderson impurity model, i.e., an interacting impurity hybridised with a non-interacting bath, an unnecessary requirement which would lead to the same problems of the DMFT iterative solution.

The expectation value per site $E$ of the Hamiltonian \eqn{Ham} at $h=0$ on the variational wavefunction \eqn{G-wf} at fixed magnetisation $m$ can be written as a functional 
of $\psi$ only, and reads~\cite{Lanata-ghost-2017}
\be
E[\psi] =
\fract{1}{V}\,\bra{\Psi_*} H_* \ket{\Psi_*} 
+ U\,\bra{\psi} n_\up\,n_\down \ket{\psi}\,,
\label{E_m}
\ee
where $V$ is the number of sites, and $n_\sigma$ the spin-$\sigma$ occupation number of the impurity. The Slater determinant $\Psi_*$ is the ground state of the non-interacting Hamiltonian 
\beal
H_* &= -\fract{t}{\sqrt{z}}\sum_{a,b=1}^3\, \sum_{<ij>\,\sigma}
\Big(R_{\sigma,a}^*\,R_{\sigma,b}^\dagga\,c^\dagger_{i\,a\sigma}\,c^\dagga_{j\,b\sigma} + 
H.c.\Big)\\
&\quad -\sum_{i,\sigma}\,\sum_{a,b=1}^3\, 
\mu_{\sigma,ab}\,\Big(c^\dagger_{i\,a\sigma}c^\dagga_{i\,b\sigma} -
\Delta_{\sigma,ab}\Big)\,,
\label{H_*}
\eal
where $z\to\infty$ is the Bethe lattice coordination number. 
The Lagrange multipliers $\mu_{\sigma,ab}$ in \eqn{H_*} enforce the constraint in Eq. (\ref{GW_constraints_2}), while the parameters $R_{\sigma,a}$ are defined in Eq.~\eqn{wavefunction_Rf}.
 The energy functional $E[\psi]$ in \eqn{E_m} must be minimised with respect to $\psi$ subject to the constraints fixing the density to 1 and $m$ to the desired value $1-2\nph$, i.e., 
\be
\label{constraints_psi}
\bra{\psi} n_\up+n_\down\ket{\psi}=1\,,\quad
\bra{\psi} n_\up-n_\down\ket{\psi}=1-2\nph\,.
\ee
Before considering the case relevant for PES, we briefly mention, in the next section, how the Mott transition in the absence of a magnetic field occurs within this variational scheme~\cite{Lanata-ghost-2017}.

\subsection{Paramagnetic Mott transition at $h=0$.}
\label{para_case}

At $h=0$ the model recovers full $SU(2)$ spin symmetry, so that, e.g., the 
impurity model wavefunction $\ket{\psi}$ can be taken as eigenstate of the 
total spin $\mathbf{S}^2$. Moreover, since the Hamiltonian
\eqn{Ham} has particle-hole symmetry, we can assume that orbital/level 1 
is the charge conjugate of 3, while orbital/level 2 is self-conjugate and thus 
is pinned at the chemical potential, which is zero. Therefore, at large $U$,  1 and 3 describe the Hubbard sidebands, while 2 the low-energy quasiparticles.

\begin{figure}
\centerline{
\includegraphics[width=0.45\textwidth]{{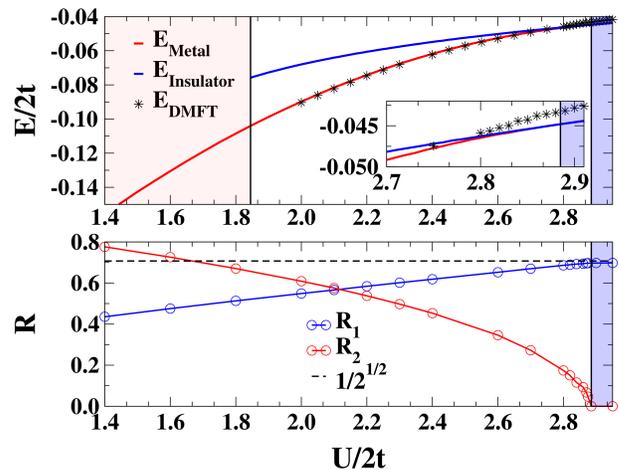}}
}
\vspace{-0.25cm}
\caption{Top panel: total energy $E$ per site vs. $U$, in units of half-bandwidth $2t$, of the stable metal and metastable insulator solutions, red and blue lines, respectively. The starred symbols are obtained by DMFT using exact diagonalization as impurity solver with 7 bath levels instead of the 3 as in the variational g-GW. In the light red region only the metallic solution exists, while in the blue one the system is a pure Mott insulator.
Bottom panel: renormalisation parameters $R_1=R_3$ and $R_2$. We note that the Mott transition is signalled by the vanishing of $R_2$, while $R_1=R_3$ approach the value $\sqrt{1/2}$.}
\label{E_GS}
\end{figure}

Minimizing the variational energy in Eq.~\eqn{E_m} with respect to 
$\psi(\Gamma,\overline{\gamma})$ we find the results shown in figure \ref{E_GS}. In agreement with \cite{Lanata-ghost-2017}, the g-GW with $N=3$ displays for $U\in[U_{c1},U_{c2}]$ the metal-insulator coexistence found in DMFT.
The values of the spinodal points, $U_{c1}/2t\simeq1.85$ and $U_{c2}/2t\simeq2.88$, are slightly underestimated with respect to DMFT, $U_{c1}/2t\simeq2.39$ 
and $U_{c2}/2t\simeq 2.94$ \cite{Hallberg2004,Werner2007,BullaPRL1999,PuPRB2001}.

\begin{figure}
\centerline{
\includegraphics[width=0.47\textwidth]{{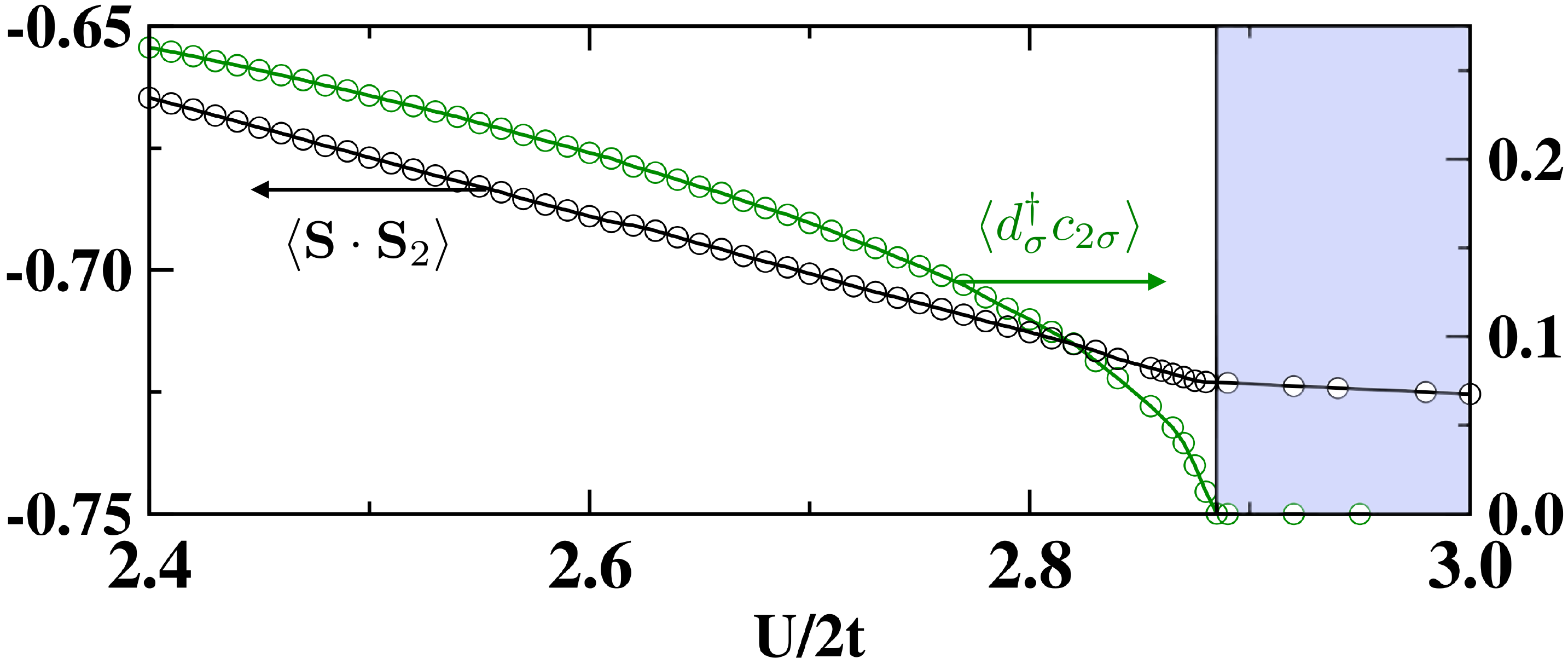}}}
\vspace{-0.25cm}
\caption{Expectation value 
$\langle\psi|\mathbf{S}\cdot\mathbf{S}_2|\psi\rangle$ of the spin correlation between the impurity and the level 2. Note that $\langle\psi|\mathbf{S}\cdot\mathbf{S}_2|\psi\rangle$ increases in absolute value with $U$, and is continuous across the Mott transition, despite the vanishing expectation value $\langle\psi|d^\dagger_\sigma\,c_{2\sigma}|\psi\rangle$ of the hybridisation, also shown.}
\label{SS}
\end{figure}

Approaching the Mott transition from the metallic side
$R_{1\sigma}=R_{3\sigma}\to 1/\sqrt{2}$ and 
$R_{2\sigma}\to 0$~\cite{Lanata-ghost-2017}, see Fig.~\ref{E_GS}.
 Just as in DMFT, the Mott transition is signalled by the vanishing hybridisation between the impurity and the bath level at the Fermi energy, $Q_{2}=\langle d^\dagger_\sigma\,c^\dagga_{2\sigma}\rangle\to 0$, 
 see Fig.~\ref{SS}. In the insulating phase, the wavefunction $|\psi\rangle$ thus factorises into a spin-1/2 wavefunction $|\phi_\sigma\rangle$ for the impurity plus levels 1 and 3, with the spin mostly localised on the impurity since 1 and 3 are far from the chemical potential, and a spin-1/2 wavefunction $|\varphi_\sigma\rangle$ of the decoupled singly-occupied level 2. However, because of equations \eqn{GW_constraints_2} and \eqn{wavefunction_Rf}, 
both of which determine the Hamiltonian $H_*$ in \eqn{H_*} and thus its ground state, the variational wavefunction with lowest energy is actually the singlet combination 
$
|\psi\rangle = (|\phi_\up\rangle\!\times\!|\varphi_\down\rangle - |\phi_\down\rangle\!\times\!|\varphi_\up\rangle)/\sqrt{2}
$
lying $\sim J/8=t^2/2U$ below the triplet.   
In other words, despite level 2 is not hybridised with the impurity, it remains entangled with the latter in the optimised wavefunction, as evident from Fig.~\ref{SS} where we plot, as a function of $U$ around the Mott transition, the expectation value $\bra{\psi}\!\mathbf{S}\cdot\mathbf{S}_2\!\ket{\psi}$, where $\mathbf{S}$ and 
$\mathbf{S}_2$ are the spin operators of the impurity and the level 2, respectively. 
We note that this quantity is continuous across the Mott transition and approaches the spin-singlet limit $-3/4$ at large $U$. 

The above result naturally suggests how to construct a good trial insulating wavefunction with fixed impurity magnetisation $m$: 
\be
|\psi\rangle = \cos\theta\,|\phi_\up\rangle\!\times\!|\varphi_\down\rangle - \sin\theta\,|\phi_\down\rangle\!\times\!|\varphi_\up\rangle,
\ee 
with $\cos2\theta\simeq m$. This wavefunction evidently describes a pure state that cannot be interpreted as the ground state of an Anderson impurity model. 
Indeed, the spin entanglement between $\ket{\phi_\sigma}$ and 
$\ket{\varphi_\sigma}$ can only be rationalised through an effective antiferromagnetic exchange between level 2 and the remaining sites of the impurity model, which is not a one-body potential.

\begin{figure}
\begin{center}
\includegraphics[width=0.52\textwidth]{{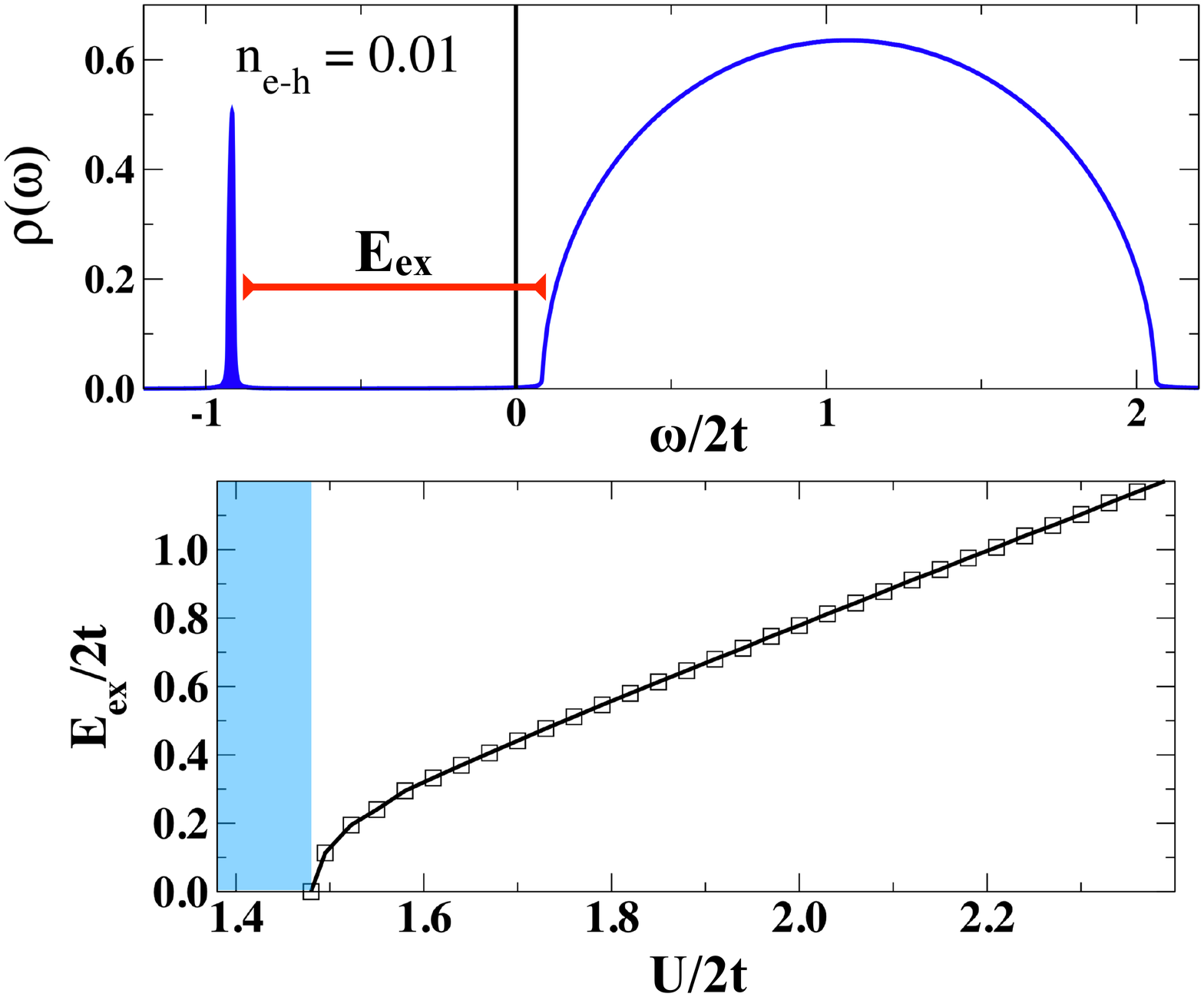}}
\caption{\label{DOS-Exc}Top panel: density of states (DOS) of an electron in the conduction band at $U/2t=2.4$ with a density $\nph=0.05$ of photoexcited e-h pairs. The narrow peak at negative frequency, the occupied side of the spectrum, corresponds to the filled exciton state, whereas the conduction band at positive frequency is empty. Such DOS thus describes an exciton gas, and allows extracting a exciton binding energy 
$E_\text{ex}$ through the 
distance of the exciton peak from the bottom of the conduction band, which is shown as function of 
$U/2t$ in the bottom panel. Note that a finite $E_\text{ex}$ requires 
$U/2t$ above a threshold $\simeq 1.5$. The region below that threshold, 
in light blue, is therefore not representative of PES.}
\end{center}
\end{figure}

\section{Exciton Mott transition in photoexcited semiconductors}
\label{Mott_PES}

We now move to study the phase diagram of the Hamiltonian \eqn{Ham} as function of its parameters, i.e., the magnetisation 
$m$ and interaction strength $U$, with the half-bandwidth $2t$ the unit of energy. However, to make contact with experiments on photoexcited semiconductors, we need to translate both of them into physical properties that characterise those systems. From Eq.~\eqn{mapping-PES} it follows that $m$ is simply related to the number $\nph$ of photoexcited e-h pairs through $m=1-2\nph$. 
On the contrary, the short-range Hubbard $U$ is not directly related to any 
physical property of PES, since it is just meant to mimic the role of the long-range Coulomb interaction in binding electrons and holes into excitons. This implies that the model Hamiltonian \eqn{Ham} must have in common with PES the existence of exciton states. We shall thus use the corresponding binding energy $E_\text{ex}$ as the other state variable besides $\nph$ to characterise the phase diagram. 

In the present case of a semicircular tight-binding DOS, whose square root behaviour at the edges reproduces that at the bottom or top of a band in three dimensions, the on-site interaction $U$ must exceed a threshold to produce a bound state, unlike Coulomb repulsion. Its binding energy $E_\text{ex}$ 
can be calculated exactly solving the problem of a single e-h pair. However, for consistency with the calculation at $\nph>0$, we determine $E_\text{ex}$ by the variational optimisation in the limit 
$\nph\to 0$. In the top panel of Fig.~\ref{DOS-Exc} we show for the model Hamiltonian \eqn{Ham} 
the single particle DOS of a spin down electron in the Mott insulator at almost full polarisation, $m=0.98$, which, through 
Eq.~\eqn{PH_transf}, corresponds to the DOS of an electron in the conduction band within the EG phase at $\nph=0.01$, as well as, since we assumed particle-hole symmetry, to the DOS of a hole in the valence band. We note at positive energies, i.e., above the chemical potential of the quasi stationary local equilibrium state 
in Fig.~\ref{toy_model}(c), the continuum of empty states in the conduction band, and at negative energy a very narrow peak that accommodates all the $\nph=0.01$ density of photoexcited electrons. This peak evidently corresponds 
to the exciton, and its distance from the bottom of the conduction band defines the binding energy 
$E_\text{ex}$, whose dependence on $U$ is shown in the bottom panel 
of the same figure ~\ref{DOS-Exc}. Only when $E_\text{ex}>0$, 
i.e., $U/2t > U_\text{ex}/2t\simeq 1.5$, the model Hamiltonian \eqn{Ham} can be used to describe the physics of PES. In the language of the repulsive Hubbard model \eqn{Ham}, 
the exciton peak below the chemical potential and the broad continuum above  translate into the lower and upper Hubbard bands, respectively, while 
the threshold value $U_\text{ex}$ is actually the limit of the Mott insulator 
spinodal value of interaction, so-called $U_{c1}$ \cite{ReviewDMFT}, when the magnetisation $m\to 1$. 

\begin{figure}
\centerline{
\includegraphics[width=0.45\textwidth]{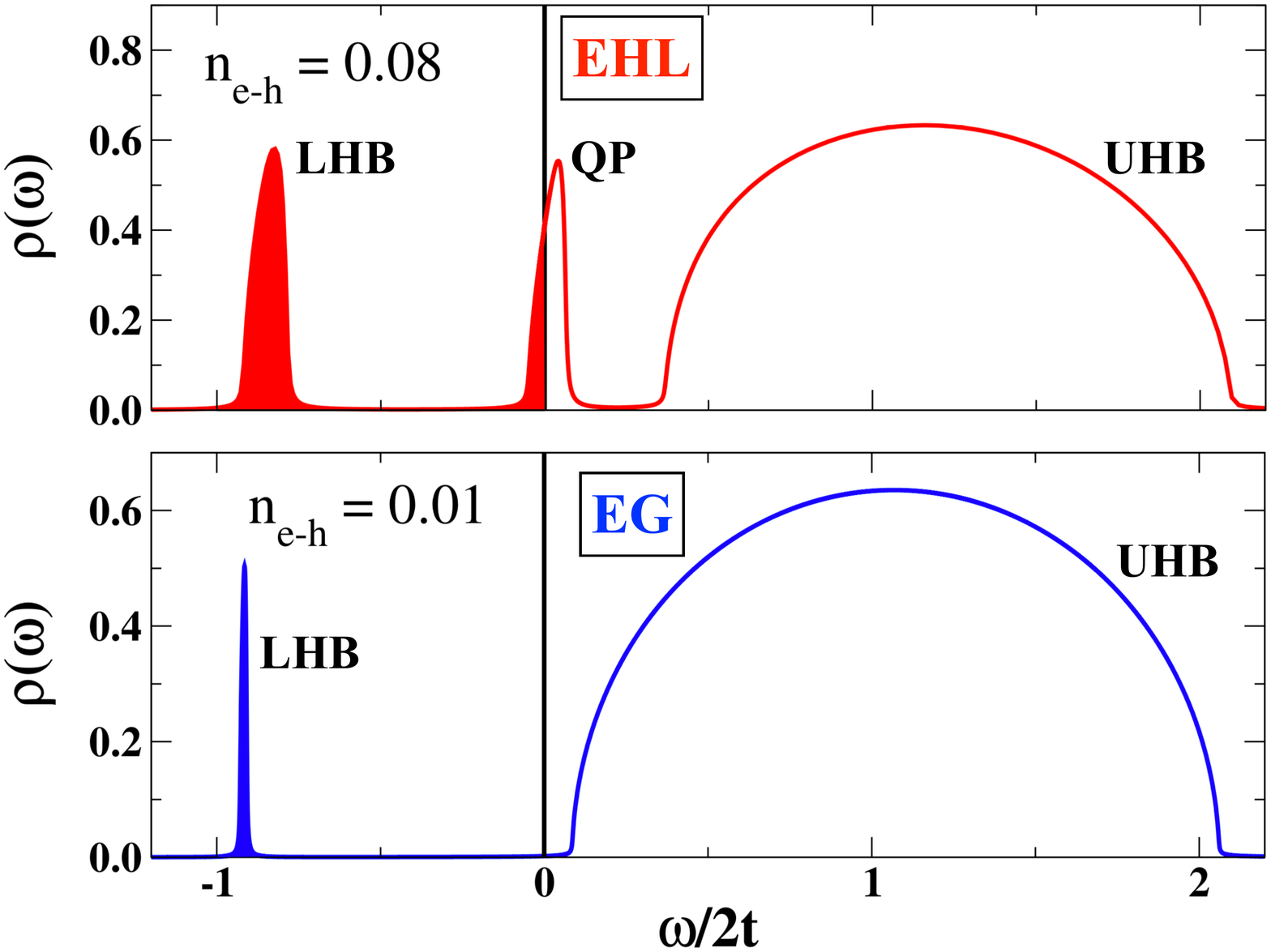}}
\vspace{-0.1cm}
\caption{ Density of states (DOS) for a spin down particle corresponding to 
an electron in the conduction band, as well as a hole in the valence one.  
Top: DOS in the metallic EHL solution at $U/2t=2.2$, 
which corresponds to $E_\text{ex}/2t\simeq 0.9$, and $\nph=0.08$. Bottom: same quantity in the EG Mott insulator at the same value of $U$ but 
smaller $\nph=0.01$. 
In the language of the repulsive Hubbard model, the features associated with 
lower and upper Hubbard bands are indicated by LHB and UHB, respectively.
Note that they both survive also in the correlated metal phase, top panel, despite the emergence of quasiparticles narrowly peaked at the Fermi energy.}
\label{DOS}
\end{figure}
Before discussing in detail the phase diagram, it is worth highlighting the 
properties that characterise the EHL as opposed to the EG. 
In Fig.~\ref{DOS} we show the DOSs of an electron in the conduction band, or a hole in the valence one, in a representative EHL solution at 
$E_\text{ex}/2t\simeq 0.9$ and $\nph=0.08$, top panel, 
in contrast to a representative EG one at the same 
$E_\text{ex}/2t\simeq 0.9$ but at smaller $\nph=0.01$, bottom panel.  
We note in the EHL phase, top panel, still clearly visible Hubbard bands, the lower one
(LHB) corresponding to the exciton, and the upper (UHB) to the incoherent 
contribution of the conduction, for the electron, and valence, for the hole, 
bands. In addition, a narrow quasiparticle peak emerges at the 
chemical potential, which distinguishes the EHL DOS in the top panel from the EG one in the bottom one. The finite gap between the quasiparticle peak and the UHB is most likely an artefact of our using a three level bath. We expect that further bath levels would fill that gap by small spectral weight. Nonetheless, the correlated metal 
feature of a coherent quasiparticle peak distinct from the UHB incoherent background should persist. \\
Our ground state calculation does not allow accessing 
the two-particle response functions associated to the optical absorption and  
luminescence. However, the DOS shown in the top panel of Fig.~\ref{DOS} suggests that the optical spectrum of the EHL should still display exciton signatures, which have been indeed observed experimentally \cite{Grivickas-PRL2003,Suzuki-PRL2012,Sekiguchi-PRL2017}, and predicted theoretically \cite{Mahan_67}. Moreover, the narrow width of the quasiparticle peak suggests an anomalous strongly correlated EHL metal, also not in disagreement with experiments \cite{Sekiguchi-PRL2017}.   \\

\begin{figure}
\centerline{
\includegraphics[width=0.5\textwidth]{{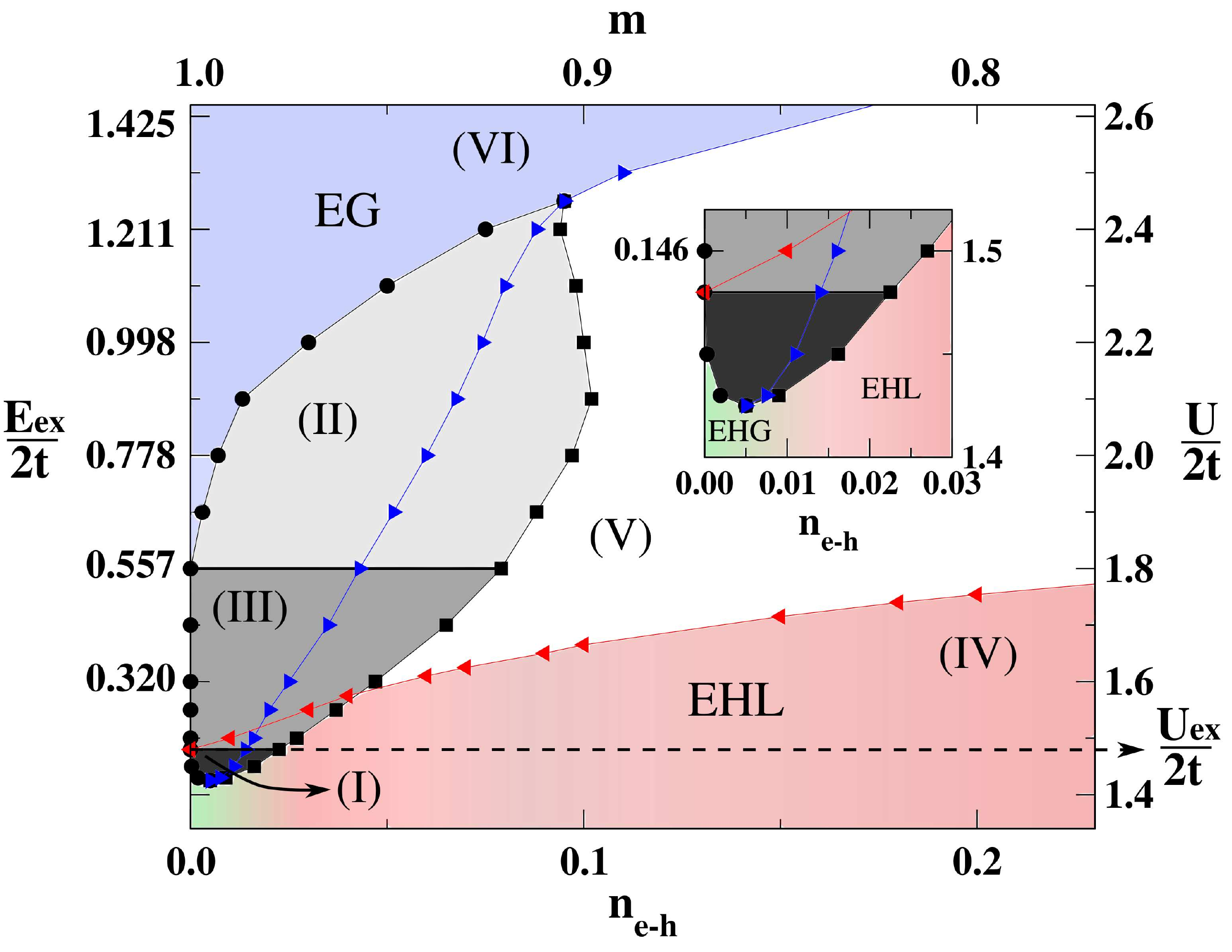}}\hspace{0.5cm}
}
\vspace{-0.15cm}
\caption{Phase diagram as function of the exciton binding energy $E_\text{ex}$ and the density $\nph$ of photoexcited electron-hole pairs, left $y$ and bottom $x$ axes, respectively, or, equivalently, 
$U/2t$ and magnetisation $m=1-2\nph$, right $y$ and top $x$ axes, respectively. In red we plot the insulator spinodal line, while in blue the metal one. The different regions, labeled from (I) to (VI) are discussed in the text. The inset shows a zoom of the phase-diagram around region (I).}
\label{phase_diagram}
\end{figure}

Let us move now to discuss the phase diagram, which is shown in Fig.~\ref{phase_diagram} 
as function of the exciton binding energy $E_\text{ex}$ and the density 
of photoexcited e-h pairs $\nph$. For completeness, we also show the values of magnetisation $m$ 
and interaction $U$ that corresponds to $\nph$ and $E_\text{ex}$ in the repulsive Hubbard model 
\eqn{Ham}. 
The red curve in that figure corresponds to the spinodal line above which an EG solution exists, which is 
also the spinodal $U_{c1}(m)$ of the Mott insulator \cite{Kotliar1999} in the repulsive model \eqn{Ham}, and becomes for $m\to 1$ the threshold value $U_\text{ex}$ in Fig.~\ref{DOS-Exc}. 
On the contrary, the blue curve is the spinodal line above which the EHL becomes unstable, and thus only the EG exists, which is the metal spinodal $U_{c2}(m)$ in the repulsive model \eqn{Ham}.   

\begin{figure}
\centerline{
\includegraphics[width=0.5\textwidth]{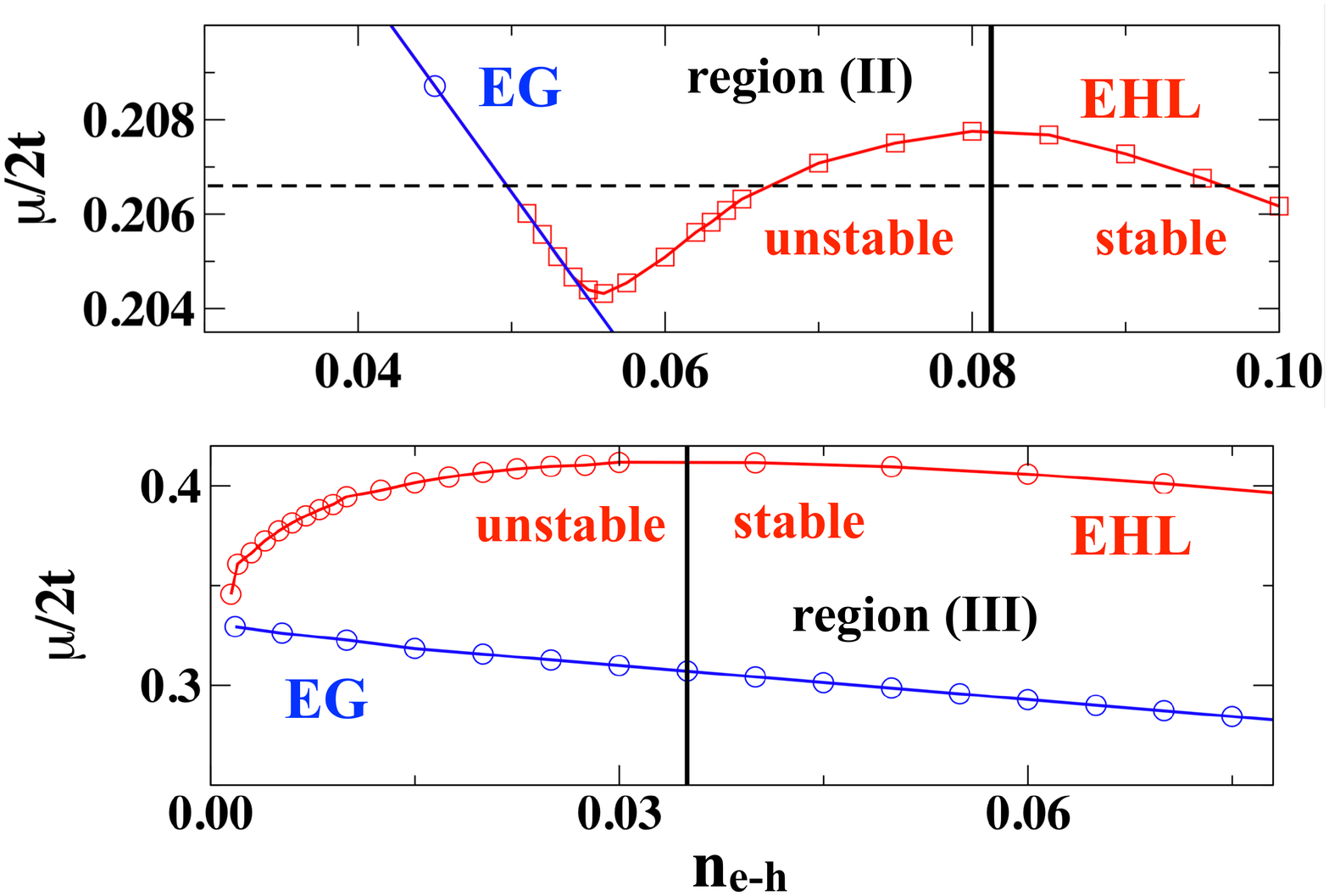}}
\vspace{-0.3cm}
\caption{Top panel: 
phase separation at $E_\text{ex}/2t\simeq 1$, region (II) in 
Fig.~\ref{phase_diagram}. We plot the chemical potential extracted by $E$ vs. $\nph$ for both 
EHL (red line) and EG (blue line) solutions. The vertical line indicate the EHL spinodal value of $\nph$, below which this phase is not stable anymore. The dotted line corresponds to the common tangent construction, which implies that for $\nph\in [0.05,0.095]$ the system phase separates into an EG at $\nph=0.05$ and 
an EHL at $\nph=0.095$.\\
Bottom panel: bistability at $E_\text{ex}/2t\simeq 1$, region (III) in Fig.~\ref{phase_diagram}. Here the behaviour of the chemical potential of the EHL (red line) and EG (blue line) solutions does not allow 
a common tangent construction. }
\label{phase-separation}
\end{figure}

Considering instead the different phases in Fig.~\ref{phase_diagram}, in the dark region (I) for very small $\nph$, 
which is zoomed in the inset and, strictly speaking, is not pertinent to PES since $U<U_\text{ex}$, we find phase separation between two metallic phases, one liquid (EHL) at larger 
$\nph$, and the other gaseous (EHG), at smaller $\nph$. The two phases merge at a second order critical point. \\
In the light grey region (II) above $E_\text{ex}/2t\simeq 0.557$, now relevant to PES, we instead find phase 
separation between an EG and an EHL. In the top panel of Fig.~\ref{phase-separation} we plot the chemical potential obtained from the energy per 
site $E$ and $\nph$ through $\mu=-\partial E/\partial \nph$, together with the common tangent construction. It follows that for $\nph\in [0.05,0.095]$ the system phase separates into an EG at low density $\nph=0.05$ and 
an EHL at larger density $\nph=0.095$.\\
On the contrary, within the intermediate region (III), dark grey in the phase diagram Fig.~\ref{phase_diagram}, we could not make a common-tangent construction, see bottom panel of Fig.~\ref{phase-separation}.
This may suggest a bistable behaviour with a sudden transformation of the EG into the EHL, which has been invoked \cite{Shimano-bistable} to explain some experimental evidences \cite{Sekiguchi-PRL2017}, even though we cannot exclude 
a numerical artefact since the variational optimisation is hard at very low $\nph$.  

In region (IV), below $U_{c1}(m)$, only the EHL is stable. On the contrary, in region (V), white in the figure, we find coexistence between a stable EHL and a metastable EG. Finally, in region (VI), blue in the figure, only the EG is stable. We find that the transition line separating (V) and (VI) has a second order character. We note that 
in the language of the repulsive Hubbard model \eqn{Ham}, this second order line persists down to $m=0$, while, for the reasons outlined in 
Sec.~\ref{method}, DMFT calculations find a continuous transition only at $m=0$, and a discontinuous one at any $m\not=0$ \cite{GeorgesPRB1994,Massimo02,HewsonPRB2007,Zhu_2017}.   
 
The phase diagram in Fig.~\ref{phase_diagram} has been obtained at zero temperature forcing all symmetries of the Hamiltonian \eqn{Ham}, and should  be representative of that above the exciton condensation temperature with the 
caveat that the character of all transition lines, and the size of 
the stability domain of each phase might change when accounting for 
entropy effects.  
Let us therefore discuss what we expect might change at finite temperature.  
Since the exciton peak in Fig.~\ref{DOS} presumably carries more entropy than 
the quasi-particle peak \cite{ReviewDMFT}, it is most likely that the second order line separating regions (V) and (VI) transforms into a first order transition accompanied by phase separation, i.e., a gradual crossover as $\nph$ increases from the EG phase to the EHL one that simply extends region (II) to higher values of $\nph$. 

On the contrary, we cannot exclude that, if the bistability in region 
(III) is true and not due to numerical issues, it might survive at finite temperature, nor that, at the border between regions (I) and (III), a discontinuous exciton Mott transition appears besides the gaseous-liquid transition $\text{EHG}\to\text{EHL}$.

\section{Conclusions}
\label{Conclusions}

Despite its extreme simplicity, the spin-polarised half-filled repulsive Hubbard model shows the rather rich phase diagram in Fig.~\ref{phase_diagram}
with phase transitions that, upon translation in the language 
of semiconductors at finite density of photoexcited electron-hole pairs, bear strong similarities with the transitions from an exciton gas to an electron-hole liquid observed in those systems, among the 
earliest known realisations of Mott transitions. In particular, 
taking into account finite temperature entropy effects not included in our calculation, the phase diagram in Fig.~\ref{phase_diagram} 
encompasses a first order exciton Mott transition that almost everywhere is accompanied by phase separation, thus implying a continuous transformation 
of the exciton gas into the electron-hole liquid. However, within a small region in the phase diagram at low exciton binding energy, 
we also find bistability, which would correspond to a sudden transformation of the exciton gas into the electron-hole liquid without phase-separation. 
We mentioned that experimentally there are both evidences of 
continuos exciton Mott transitions, i.e., phase separation 
\cite{Kappei-PRL2005,Rossbach-PRB2014,Kirsanke-PRB2016,Chernikov-NaturePhotonics2015},  
as well as of discontinuous ones \cite{Wolfe-PRL1986,Wolfe-PRB1995,Amo-2007,Stern-PRL2008,Sekiguchi-PRL2017,Shimano-bistable,Bataller-NanoLett2019}. The discriminant parameter might well be the exciton 
binding energy $E_\text{ex}$, 
as we do find, since a discontinuous transition is 
mostly observed in bulk semiconductors, while a continuous one in 
confined geometries, like quantum wells, where $E_\text{ex}$ is 
supposedly larger. 
Transition metal dichalcogenides seem to constitute an exception to this rule, 
since, despite their large exciton binding energy, they have been shown to display either a continuous transition \cite{Chernikov-NaturePhotonics2015} 
after an ultrashort laser pulse, or a discontinuous one \cite{Bataller-NanoLett2019} under continuous-wave photoexcitation, although we cannot exclude that the different photoexcitation processes are responsible of the different outcomes. \\
We end remarking that the electron-hole liquid that we find is rather anomalous, see the single-particle density of states shown in the top panel of Fig.~\ref{DOS}, since it still displays 
excitonic signatures despite being a degenerate Fermi liquid of holes 
and electrons, at odds with the expectation that the exciton binding energy 
should vanish at the transition. Moreover, we find that the effective mass of the unbound electrons and holes is quite large, as clear from the narrow 
width of the coherent peak at the chemical potential. Both these features 
have been observed experimentally \cite{Sekiguchi-PRL2017}.

\begin{acknowledgments}

We are grateful to Nicola Lanat\`a, Adriano Amaricci and Roberto Raimondi for fruitful discussions.
This work has been supported by 
the European Union under H2020 Framework Programs, ERC Advanced Grant No. 692670 ``FIRSTORM''. 
M.C. acknowledges support from MIUR PRIN 2015 (Prot.2015C5SEJJ001) and SISSA/CNR project Superconductivity, Ferroelectricity and Magnetism in bad metals (Prot.232/2015).

\end{acknowledgments}
 

\bibliographystyle{apsrev}
\bibliography{mybiblio}

\begin{thebibliography}{65}
\expandafter\ifx\csname natexlab\endcsname\relax\def\natexlab#1{#1}\fi
\expandafter\ifx\csname bibnamefont\endcsname\relax
  \def\bibnamefont#1{#1}\fi
\expandafter\ifx\csname bibfnamefont\endcsname\relax
  \def\bibfnamefont#1{#1}\fi
\expandafter\ifx\csname citenamefont\endcsname\relax
  \def\citenamefont#1{#1}\fi
\expandafter\ifx\csname url\endcsname\relax
  \def\url#1{\texttt{#1}}\fi
\expandafter\ifx\csname urlprefix\endcsname\relax\def\urlprefix{URL }\fi
\providecommand{\bibinfo}[2]{#2}
\providecommand{\eprint}[2][]{\url{#2}}

\bibitem[{\citenamefont{Brinkman and Rice}(1973)}]{Brinkman&Rice-PRB1973}
\bibinfo{author}{\bibfnamefont{W.~F.} \bibnamefont{Brinkman}} \bibnamefont{and}
  \bibinfo{author}{\bibfnamefont{T.~M.} \bibnamefont{Rice}},
  \bibinfo{journal}{Phys. Rev. B} \textbf{\bibinfo{volume}{7}},
  \bibinfo{pages}{1508} (\bibinfo{year}{1973}),
  \urlprefix\url{https://link.aps.org/doi/10.1103/PhysRevB.7.1508}.

\bibitem[{\citenamefont{Mott}(1973)}]{Mott-book}
\bibinfo{author}{\bibfnamefont{N.~F.} \bibnamefont{Mott}},
  \bibinfo{journal}{Contemporary Physics} \textbf{\bibinfo{volume}{14}},
  \bibinfo{pages}{401} (\bibinfo{year}{1973}),
  \eprint{https://doi.org/10.1080/00107517308210764},
  \urlprefix\url{https://doi.org/10.1080/00107517308210764}.

\bibitem[{\citenamefont{Rice}(1978)}]{Rice-SSP1978}
\bibinfo{author}{\bibfnamefont{T.}~\bibnamefont{Rice}}
  (\bibinfo{publisher}{Academic Press}, \bibinfo{year}{1978}),
  vol.~\bibinfo{volume}{32} of \emph{\bibinfo{series}{Solid State Physics}},
  pp. \bibinfo{pages}{1 -- 86},
  \urlprefix\url{http://www.sciencedirect.com/science/article/pii/S0081194708604385}.

\bibitem[{\citenamefont{Mott}(1949)}]{Mott-1949}
\bibinfo{author}{\bibfnamefont{N.~F.} \bibnamefont{Mott}},
  \bibinfo{journal}{Proceedings of the Physical Society. Section A}
  \textbf{\bibinfo{volume}{62}}, \bibinfo{pages}{416} (\bibinfo{year}{1949}),
  \urlprefix\url{http://stacks.iop.org/0370-1298/62/i=7/a=303}.

\bibitem[{\citenamefont{Landau and Zeldovich}(1943)}]{Landau&Zeldovich}
\bibinfo{author}{\bibfnamefont{L.}~\bibnamefont{Landau}} \bibnamefont{and}
  \bibinfo{author}{\bibfnamefont{Y.}~\bibnamefont{Zeldovich}},
  \bibinfo{journal}{Acta Phys. Chem. URSS} \textbf{\bibinfo{volume}{18}},
  \bibinfo{pages}{194} (\bibinfo{year}{1943}).

\bibitem[{\citenamefont{Sekiguchi and Shimano}(2017)}]{Shimano-bistable}
\bibinfo{author}{\bibfnamefont{F.}~\bibnamefont{Sekiguchi}} \bibnamefont{and}
  \bibinfo{author}{\bibfnamefont{R.}~\bibnamefont{Shimano}},
  \bibinfo{journal}{Journal of the Physical Society of Japan}
  \textbf{\bibinfo{volume}{86}}, \bibinfo{pages}{103702}
  (\bibinfo{year}{2017}), \eprint{https://doi.org/10.7566/JPSJ.86.103702},
  \urlprefix\url{https://doi.org/10.7566/JPSJ.86.103702}.

\bibitem[{\citenamefont{Smith and Wolfe}(1986)}]{Wolfe-PRL1986}
\bibinfo{author}{\bibfnamefont{L.~M.} \bibnamefont{Smith}} \bibnamefont{and}
  \bibinfo{author}{\bibfnamefont{J.~P.} \bibnamefont{Wolfe}},
  \bibinfo{journal}{Phys. Rev. Lett.} \textbf{\bibinfo{volume}{57}},
  \bibinfo{pages}{2314} (\bibinfo{year}{1986}),
  \urlprefix\url{https://link.aps.org/doi/10.1103/PhysRevLett.57.2314}.

\bibitem[{\citenamefont{Simon et~al.}(1992)\citenamefont{Simon, Kirch, and
  Wolfe}}]{Wolfe-PRB1992}
\bibinfo{author}{\bibfnamefont{A.~H.} \bibnamefont{Simon}},
  \bibinfo{author}{\bibfnamefont{S.~J.} \bibnamefont{Kirch}}, \bibnamefont{and}
  \bibinfo{author}{\bibfnamefont{J.~P.} \bibnamefont{Wolfe}},
  \bibinfo{journal}{Phys. Rev. B} \textbf{\bibinfo{volume}{46}},
  \bibinfo{pages}{10098} (\bibinfo{year}{1992}),
  \urlprefix\url{https://link.aps.org/doi/10.1103/PhysRevB.46.10098}.

\bibitem[{\citenamefont{Amo et~al.}(2007)\citenamefont{Amo, Martín, Viña,
  Toropov, and Zhuravlev}}]{Amo-2007}
\bibinfo{author}{\bibfnamefont{A.}~\bibnamefont{Amo}},
  \bibinfo{author}{\bibfnamefont{M.~D.} \bibnamefont{Martín}},
  \bibinfo{author}{\bibfnamefont{L.}~\bibnamefont{Viña}},
  \bibinfo{author}{\bibfnamefont{A.~I.} \bibnamefont{Toropov}},
  \bibnamefont{and} \bibinfo{author}{\bibfnamefont{K.~S.}
  \bibnamefont{Zhuravlev}}, \bibinfo{journal}{Journal of Applied Physics}
  \textbf{\bibinfo{volume}{101}}, \bibinfo{pages}{081717}
  (\bibinfo{year}{2007}), \eprint{https://doi.org/10.1063/1.2722786},
  \urlprefix\url{https://doi.org/10.1063/1.2722786}.

\bibitem[{\citenamefont{Stern et~al.}(2008)\citenamefont{Stern, Garmider,
  Umansky, and Bar-Joseph}}]{Stern-PRL2008}
\bibinfo{author}{\bibfnamefont{M.}~\bibnamefont{Stern}},
  \bibinfo{author}{\bibfnamefont{V.}~\bibnamefont{Garmider}},
  \bibinfo{author}{\bibfnamefont{V.}~\bibnamefont{Umansky}}, \bibnamefont{and}
  \bibinfo{author}{\bibfnamefont{I.}~\bibnamefont{Bar-Joseph}},
  \bibinfo{journal}{Phys. Rev. Lett.} \textbf{\bibinfo{volume}{100}},
  \bibinfo{pages}{256402} (\bibinfo{year}{2008}),
  \urlprefix\url{https://link.aps.org/doi/10.1103/PhysRevLett.100.256402}.

\bibitem[{\citenamefont{Kappei et~al.}(2005)\citenamefont{Kappei, Szczytko,
  Morier-Genoud, and Deveaud}}]{Kappei-PRL2005}
\bibinfo{author}{\bibfnamefont{L.}~\bibnamefont{Kappei}},
  \bibinfo{author}{\bibfnamefont{J.}~\bibnamefont{Szczytko}},
  \bibinfo{author}{\bibfnamefont{F.}~\bibnamefont{Morier-Genoud}},
  \bibnamefont{and} \bibinfo{author}{\bibfnamefont{B.}~\bibnamefont{Deveaud}},
  \bibinfo{journal}{Phys. Rev. Lett.} \textbf{\bibinfo{volume}{94}},
  \bibinfo{pages}{147403} (\bibinfo{year}{2005}),
  \urlprefix\url{https://link.aps.org/doi/10.1103/PhysRevLett.94.147403}.

\bibitem[{\citenamefont{Suzuki and Shimano}(2012)}]{Suzuki-PRL2012}
\bibinfo{author}{\bibfnamefont{T.}~\bibnamefont{Suzuki}} \bibnamefont{and}
  \bibinfo{author}{\bibfnamefont{R.}~\bibnamefont{Shimano}},
  \bibinfo{journal}{Phys. Rev. Lett.} \textbf{\bibinfo{volume}{109}},
  \bibinfo{pages}{046402} (\bibinfo{year}{2012}),
  \urlprefix\url{https://link.aps.org/doi/10.1103/PhysRevLett.109.046402}.

\bibitem[{\citenamefont{Rossbach et~al.}(2014)\citenamefont{Rossbach, Levrat,
  Jacopin, Shahmohammadi, Carlin, Gani\`ere, Butt\'e, Deveaud, and
  Grandjean}}]{Rossbach-PRB2014}
\bibinfo{author}{\bibfnamefont{G.}~\bibnamefont{Rossbach}},
  \bibinfo{author}{\bibfnamefont{J.}~\bibnamefont{Levrat}},
  \bibinfo{author}{\bibfnamefont{G.}~\bibnamefont{Jacopin}},
  \bibinfo{author}{\bibfnamefont{M.}~\bibnamefont{Shahmohammadi}},
  \bibinfo{author}{\bibfnamefont{J.-F.} \bibnamefont{Carlin}},
  \bibinfo{author}{\bibfnamefont{J.-D.} \bibnamefont{Gani\`ere}},
  \bibinfo{author}{\bibfnamefont{R.}~\bibnamefont{Butt\'e}},
  \bibinfo{author}{\bibfnamefont{B.}~\bibnamefont{Deveaud}}, \bibnamefont{and}
  \bibinfo{author}{\bibfnamefont{N.}~\bibnamefont{Grandjean}},
  \bibinfo{journal}{Phys. Rev. B} \textbf{\bibinfo{volume}{90}},
  \bibinfo{pages}{201308(R)} (\bibinfo{year}{2014}),
  \urlprefix\url{https://link.aps.org/doi/10.1103/PhysRevB.90.201308}.

\bibitem[{\citenamefont{Kir\ifmmode \check{s}\else
  \v{s}\fi{}ansk\ifmmode~\dot{e}\else \.{e}\fi{}
  et~al.}(2016)\citenamefont{Kir\ifmmode \check{s}\else
  \v{s}\fi{}ansk\ifmmode~\dot{e}\else \.{e}\fi{}, Tighineanu, Daveau,
  Miguel-S\'anchez, Lodahl, and Stobbe}}]{Kirsanke-PRB2016}
\bibinfo{author}{\bibfnamefont{G.}~\bibnamefont{Kir\ifmmode \check{s}\else
  \v{s}\fi{}ansk\ifmmode~\dot{e}\else \.{e}\fi{}}},
  \bibinfo{author}{\bibfnamefont{P.}~\bibnamefont{Tighineanu}},
  \bibinfo{author}{\bibfnamefont{R.~S.} \bibnamefont{Daveau}},
  \bibinfo{author}{\bibfnamefont{J.}~\bibnamefont{Miguel-S\'anchez}},
  \bibinfo{author}{\bibfnamefont{P.}~\bibnamefont{Lodahl}}, \bibnamefont{and}
  \bibinfo{author}{\bibfnamefont{S.}~\bibnamefont{Stobbe}},
  \bibinfo{journal}{Phys. Rev. B} \textbf{\bibinfo{volume}{94}},
  \bibinfo{pages}{155438} (\bibinfo{year}{2016}),
  \urlprefix\url{https://link.aps.org/doi/10.1103/PhysRevB.94.155438}.

\bibitem[{\citenamefont{Wang et~al.}(2012)\citenamefont{Wang, Kalantar-Zadeh,
  Kis, Coleman, and Strano}}]{Wang-NatureNano2012}
\bibinfo{author}{\bibfnamefont{Q.~H.} \bibnamefont{Wang}},
  \bibinfo{author}{\bibfnamefont{K.}~\bibnamefont{Kalantar-Zadeh}},
  \bibinfo{author}{\bibfnamefont{A.}~\bibnamefont{Kis}},
  \bibinfo{author}{\bibfnamefont{J.~N.} \bibnamefont{Coleman}},
  \bibnamefont{and} \bibinfo{author}{\bibfnamefont{M.~S.}
  \bibnamefont{Strano}}, \bibinfo{journal}{Nature Nanotechnology}
  \textbf{\bibinfo{volume}{7}}, \bibinfo{pages}{699 EP }
  (\bibinfo{year}{2012}),
  \urlprefix\url{https://doi.org/10.1038/nnano.2012.193}.

\bibitem[{\citenamefont{Duan et~al.}(2015)\citenamefont{Duan, Wang, Pan, Yu,
  and Duan}}]{Duan-ChemSocRev2015}
\bibinfo{author}{\bibfnamefont{X.}~\bibnamefont{Duan}},
  \bibinfo{author}{\bibfnamefont{C.}~\bibnamefont{Wang}},
  \bibinfo{author}{\bibfnamefont{A.}~\bibnamefont{Pan}},
  \bibinfo{author}{\bibfnamefont{R.}~\bibnamefont{Yu}}, \bibnamefont{and}
  \bibinfo{author}{\bibfnamefont{X.}~\bibnamefont{Duan}},
  \bibinfo{journal}{Chem. Soc. Rev.} \textbf{\bibinfo{volume}{44}},
  \bibinfo{pages}{8859} (\bibinfo{year}{2015}),
  \urlprefix\url{http://dx.doi.org/10.1039/C5CS00507H}.

\bibitem[{\citenamefont{Xiao et~al.}(2016)\citenamefont{Xiao, Zhao, Wang, and
  Zhang}}]{Xiao-Nanoph2016}
\bibinfo{author}{\bibfnamefont{J.}~\bibnamefont{Xiao}},
  \bibinfo{author}{\bibfnamefont{M.}~\bibnamefont{Zhao}},
  \bibinfo{author}{\bibfnamefont{Y.}~\bibnamefont{Wang}}, \bibnamefont{and}
  \bibinfo{author}{\bibfnamefont{X.}~\bibnamefont{Zhang}},
  \bibinfo{journal}{Nanophotonics} \textbf{\bibinfo{volume}{6}},
  \bibinfo{pages}{1309} (\bibinfo{year}{2016}),
  \urlprefix\url{https://doi.org/10.1515/nanoph-2016-0160}.

\bibitem[{\citenamefont{Yuan et~al.}(2017)\citenamefont{Yuan, Wang, Zhu, Zhou,
  and Huang}}]{Yuan-JPCL2017}
\bibinfo{author}{\bibfnamefont{L.}~\bibnamefont{Yuan}},
  \bibinfo{author}{\bibfnamefont{T.}~\bibnamefont{Wang}},
  \bibinfo{author}{\bibfnamefont{T.}~\bibnamefont{Zhu}},
  \bibinfo{author}{\bibfnamefont{M.}~\bibnamefont{Zhou}}, \bibnamefont{and}
  \bibinfo{author}{\bibfnamefont{L.}~\bibnamefont{Huang}},
  \bibinfo{journal}{The Journal of Physical Chemistry Letters}
  \textbf{\bibinfo{volume}{8}}, \bibinfo{pages}{3371} (\bibinfo{year}{2017}),
  \urlprefix\url{https://doi.org/10.1021/acs.jpclett.7b00885}.

\bibitem[{\citenamefont{Wang et~al.}(2018)\citenamefont{Wang, Chernikov,
  Glazov, Heinz, Marie, Amand, and Urbaszek}}]{Wang-RMP2018}
\bibinfo{author}{\bibfnamefont{G.}~\bibnamefont{Wang}},
  \bibinfo{author}{\bibfnamefont{A.}~\bibnamefont{Chernikov}},
  \bibinfo{author}{\bibfnamefont{M.~M.} \bibnamefont{Glazov}},
  \bibinfo{author}{\bibfnamefont{T.~F.} \bibnamefont{Heinz}},
  \bibinfo{author}{\bibfnamefont{X.}~\bibnamefont{Marie}},
  \bibinfo{author}{\bibfnamefont{T.}~\bibnamefont{Amand}}, \bibnamefont{and}
  \bibinfo{author}{\bibfnamefont{B.}~\bibnamefont{Urbaszek}},
  \bibinfo{journal}{Rev. Mod. Phys.} \textbf{\bibinfo{volume}{90}},
  \bibinfo{pages}{021001} (\bibinfo{year}{2018}),
  \urlprefix\url{https://link.aps.org/doi/10.1103/RevModPhys.90.021001}.

\bibitem[{\citenamefont{He et~al.}(2014)\citenamefont{He, Kumar, Zhao, Wang,
  Mak, Zhao, and Shan}}]{exciton-binding-dichalcogenides-1}
\bibinfo{author}{\bibfnamefont{K.}~\bibnamefont{He}},
  \bibinfo{author}{\bibfnamefont{N.}~\bibnamefont{Kumar}},
  \bibinfo{author}{\bibfnamefont{L.}~\bibnamefont{Zhao}},
  \bibinfo{author}{\bibfnamefont{Z.}~\bibnamefont{Wang}},
  \bibinfo{author}{\bibfnamefont{K.~F.} \bibnamefont{Mak}},
  \bibinfo{author}{\bibfnamefont{H.}~\bibnamefont{Zhao}}, \bibnamefont{and}
  \bibinfo{author}{\bibfnamefont{J.}~\bibnamefont{Shan}},
  \bibinfo{journal}{Phys. Rev. Lett.} \textbf{\bibinfo{volume}{113}},
  \bibinfo{pages}{026803} (\bibinfo{year}{2014}),
  \urlprefix\url{https://link.aps.org/doi/10.1103/PhysRevLett.113.026803}.

\bibitem[{\citenamefont{Ugeda et~al.}(2014)\citenamefont{Ugeda, Bradley, Shi,
  da~Jornada, Zhang, Qiu, Ruan, Mo, Hussain, Shen
  et~al.}}]{exciton-binding-dichalcogenides-2}
\bibinfo{author}{\bibfnamefont{M.~M.} \bibnamefont{Ugeda}},
  \bibinfo{author}{\bibfnamefont{A.~J.} \bibnamefont{Bradley}},
  \bibinfo{author}{\bibfnamefont{S.-F.} \bibnamefont{Shi}},
  \bibinfo{author}{\bibfnamefont{F.~H.} \bibnamefont{da~Jornada}},
  \bibinfo{author}{\bibfnamefont{Y.}~\bibnamefont{Zhang}},
  \bibinfo{author}{\bibfnamefont{D.~Y.} \bibnamefont{Qiu}},
  \bibinfo{author}{\bibfnamefont{W.}~\bibnamefont{Ruan}},
  \bibinfo{author}{\bibfnamefont{S.-K.} \bibnamefont{Mo}},
  \bibinfo{author}{\bibfnamefont{Z.}~\bibnamefont{Hussain}},
  \bibinfo{author}{\bibfnamefont{Z.-X.} \bibnamefont{Shen}},
  \bibnamefont{et~al.}, \bibinfo{journal}{Nature Materials}
  \textbf{\bibinfo{volume}{13}}, \bibinfo{pages}{1091 EP }
  (\bibinfo{year}{2014}), \urlprefix\url{https://doi.org/10.1038/nmat4061}.

\bibitem[{\citenamefont{Park et~al.}(2018)\citenamefont{Park, Mutz, Schultz,
  Blumstengel, Han, Aljarb, Li, List-Kratochvil, Amsalem, and
  Koch}}]{exciton-binding-dichalcogenides-3}
\bibinfo{author}{\bibfnamefont{S.}~\bibnamefont{Park}},
  \bibinfo{author}{\bibfnamefont{N.}~\bibnamefont{Mutz}},
  \bibinfo{author}{\bibfnamefont{T.}~\bibnamefont{Schultz}},
  \bibinfo{author}{\bibfnamefont{S.}~\bibnamefont{Blumstengel}},
  \bibinfo{author}{\bibfnamefont{A.}~\bibnamefont{Han}},
  \bibinfo{author}{\bibfnamefont{A.}~\bibnamefont{Aljarb}},
  \bibinfo{author}{\bibfnamefont{L.-J.} \bibnamefont{Li}},
  \bibinfo{author}{\bibfnamefont{E.~J.~W.} \bibnamefont{List-Kratochvil}},
  \bibinfo{author}{\bibfnamefont{P.}~\bibnamefont{Amsalem}}, \bibnamefont{and}
  \bibinfo{author}{\bibfnamefont{N.}~\bibnamefont{Koch}}, \bibinfo{journal}{2D
  Materials} \textbf{\bibinfo{volume}{5}}, \bibinfo{pages}{025003}
  (\bibinfo{year}{2018}),
  \urlprefix\url{https://doi.org/10.1088%2F2053-1583%2Faaa4ca}.

\bibitem[{\citenamefont{Mueller and Malic}(2018)}]{Mueller-npj2018}
\bibinfo{author}{\bibfnamefont{T.}~\bibnamefont{Mueller}} \bibnamefont{and}
  \bibinfo{author}{\bibfnamefont{E.}~\bibnamefont{Malic}},
  \bibinfo{journal}{npj 2D Materials and Applications}
  \textbf{\bibinfo{volume}{2}}, \bibinfo{pages}{29} (\bibinfo{year}{2018}),
  \urlprefix\url{https://doi.org/10.1038/s41699-018-0074-2}.

\bibitem[{\citenamefont{Mak et~al.}(2012)\citenamefont{Mak, He, Shan, and
  Heinz}}]{Mak-NatNanotec2012}
\bibinfo{author}{\bibfnamefont{K.~F.} \bibnamefont{Mak}},
  \bibinfo{author}{\bibfnamefont{K.}~\bibnamefont{He}},
  \bibinfo{author}{\bibfnamefont{J.}~\bibnamefont{Shan}}, \bibnamefont{and}
  \bibinfo{author}{\bibfnamefont{T.~F.} \bibnamefont{Heinz}},
  \bibinfo{journal}{Nature Nanotechnology} \textbf{\bibinfo{volume}{7}},
  \bibinfo{pages}{494 EP } (\bibinfo{year}{2012}),
  \urlprefix\url{https://doi.org/10.1038/nnano.2012.96}.

\bibitem[{\citenamefont{Li et~al.}(2014)\citenamefont{Li, Ludwig, Low,
  Chernikov, Cui, Arefe, Kim, van~der Zande, Rigosi, Hill et~al.}}]{Li-PRL2014}
\bibinfo{author}{\bibfnamefont{Y.}~\bibnamefont{Li}},
  \bibinfo{author}{\bibfnamefont{J.}~\bibnamefont{Ludwig}},
  \bibinfo{author}{\bibfnamefont{T.}~\bibnamefont{Low}},
  \bibinfo{author}{\bibfnamefont{A.}~\bibnamefont{Chernikov}},
  \bibinfo{author}{\bibfnamefont{X.}~\bibnamefont{Cui}},
  \bibinfo{author}{\bibfnamefont{G.}~\bibnamefont{Arefe}},
  \bibinfo{author}{\bibfnamefont{Y.~D.} \bibnamefont{Kim}},
  \bibinfo{author}{\bibfnamefont{A.~M.} \bibnamefont{van~der Zande}},
  \bibinfo{author}{\bibfnamefont{A.}~\bibnamefont{Rigosi}},
  \bibinfo{author}{\bibfnamefont{H.~M.} \bibnamefont{Hill}},
  \bibnamefont{et~al.}, \bibinfo{journal}{Phys. Rev. Lett.}
  \textbf{\bibinfo{volume}{113}}, \bibinfo{pages}{266804}
  (\bibinfo{year}{2014}),
  \urlprefix\url{https://link.aps.org/doi/10.1103/PhysRevLett.113.266804}.

\bibitem[{\citenamefont{Shimazaki et~al.}(2015)\citenamefont{Shimazaki,
  Yamamoto, Borzenets, Watanabe, Taniguchi, and
  Tarucha}}]{Shimazaki-NatPhys2015}
\bibinfo{author}{\bibfnamefont{Y.}~\bibnamefont{Shimazaki}},
  \bibinfo{author}{\bibfnamefont{M.}~\bibnamefont{Yamamoto}},
  \bibinfo{author}{\bibfnamefont{I.~V.} \bibnamefont{Borzenets}},
  \bibinfo{author}{\bibfnamefont{K.}~\bibnamefont{Watanabe}},
  \bibinfo{author}{\bibfnamefont{T.}~\bibnamefont{Taniguchi}},
  \bibnamefont{and} \bibinfo{author}{\bibfnamefont{S.}~\bibnamefont{Tarucha}},
  \bibinfo{journal}{Nature Physics} \textbf{\bibinfo{volume}{11}},
  \bibinfo{pages}{1032 EP } (\bibinfo{year}{2015}),
  \urlprefix\url{https://doi.org/10.1038/nphys3551}.

\bibitem[{\citenamefont{Jin et~al.}(2018)\citenamefont{Jin, Kim, Utama, Regan,
  Kleemann, Cai, Shen, Shinner, Sengupta, Watanabe et~al.}}]{Jin-Science2018}
\bibinfo{author}{\bibfnamefont{C.}~\bibnamefont{Jin}},
  \bibinfo{author}{\bibfnamefont{J.}~\bibnamefont{Kim}},
  \bibinfo{author}{\bibfnamefont{M.~I.~B.} \bibnamefont{Utama}},
  \bibinfo{author}{\bibfnamefont{E.~C.} \bibnamefont{Regan}},
  \bibinfo{author}{\bibfnamefont{H.}~\bibnamefont{Kleemann}},
  \bibinfo{author}{\bibfnamefont{H.}~\bibnamefont{Cai}},
  \bibinfo{author}{\bibfnamefont{Y.}~\bibnamefont{Shen}},
  \bibinfo{author}{\bibfnamefont{M.~J.} \bibnamefont{Shinner}},
  \bibinfo{author}{\bibfnamefont{A.}~\bibnamefont{Sengupta}},
  \bibinfo{author}{\bibfnamefont{K.}~\bibnamefont{Watanabe}},
  \bibnamefont{et~al.}, \bibinfo{journal}{Science}
  \textbf{\bibinfo{volume}{360}}, \bibinfo{pages}{893} (\bibinfo{year}{2018}),
  ISSN \bibinfo{issn}{0036-8075},
  \eprint{https://science.sciencemag.org/content/360/6391/893.full.pdf},
  \urlprefix\url{https://science.sciencemag.org/content/360/6391/893}.

\bibitem[{\citenamefont{Chernikov et~al.}(2015)\citenamefont{Chernikov,
  Ruppert, Hill, Rigosi, and Heinz}}]{Chernikov-NaturePhotonics2015}
\bibinfo{author}{\bibfnamefont{A.}~\bibnamefont{Chernikov}},
  \bibinfo{author}{\bibfnamefont{C.}~\bibnamefont{Ruppert}},
  \bibinfo{author}{\bibfnamefont{H.~M.} \bibnamefont{Hill}},
  \bibinfo{author}{\bibfnamefont{A.~F.} \bibnamefont{Rigosi}},
  \bibnamefont{and} \bibinfo{author}{\bibfnamefont{T.~F.} \bibnamefont{Heinz}},
  \bibinfo{journal}{Nature Photonics} \textbf{\bibinfo{volume}{9}},
  \bibinfo{pages}{466 EP } (\bibinfo{year}{2015}),
  \urlprefix\url{http://dx.doi.org/10.1038/nphoton.2015.104}.

\bibitem[{\citenamefont{Pogna et~al.}(2016)\citenamefont{Pogna, Marsili,
  De~Fazio, Dal~Conte, Manzoni, Sangalli, Yoon, Lombardo, Ferrari, Marini
  et~al.}}]{Pogna-ACSNano2016}
\bibinfo{author}{\bibfnamefont{E.~A.~A.} \bibnamefont{Pogna}},
  \bibinfo{author}{\bibfnamefont{M.}~\bibnamefont{Marsili}},
  \bibinfo{author}{\bibfnamefont{D.}~\bibnamefont{De~Fazio}},
  \bibinfo{author}{\bibfnamefont{S.}~\bibnamefont{Dal~Conte}},
  \bibinfo{author}{\bibfnamefont{C.}~\bibnamefont{Manzoni}},
  \bibinfo{author}{\bibfnamefont{D.}~\bibnamefont{Sangalli}},
  \bibinfo{author}{\bibfnamefont{D.}~\bibnamefont{Yoon}},
  \bibinfo{author}{\bibfnamefont{A.}~\bibnamefont{Lombardo}},
  \bibinfo{author}{\bibfnamefont{A.~C.} \bibnamefont{Ferrari}},
  \bibinfo{author}{\bibfnamefont{A.}~\bibnamefont{Marini}},
  \bibnamefont{et~al.}, \bibinfo{journal}{ACS Nano}
  \textbf{\bibinfo{volume}{10}}, \bibinfo{pages}{1182} (\bibinfo{year}{2016}),
  \urlprefix\url{https://doi.org/10.1021/acsnano.5b06488}.

\bibitem[{\citenamefont{Steinleitner et~al.}(2017)\citenamefont{Steinleitner,
  Merkl, Nagler, Mornhinweg, Schüller, Korn, Chernikov, and
  Huber}}]{Steinleitner-NanoLett2017}
\bibinfo{author}{\bibfnamefont{P.}~\bibnamefont{Steinleitner}},
  \bibinfo{author}{\bibfnamefont{P.}~\bibnamefont{Merkl}},
  \bibinfo{author}{\bibfnamefont{P.}~\bibnamefont{Nagler}},
  \bibinfo{author}{\bibfnamefont{J.}~\bibnamefont{Mornhinweg}},
  \bibinfo{author}{\bibfnamefont{C.}~\bibnamefont{Schüller}},
  \bibinfo{author}{\bibfnamefont{T.}~\bibnamefont{Korn}},
  \bibinfo{author}{\bibfnamefont{A.}~\bibnamefont{Chernikov}},
  \bibnamefont{and} \bibinfo{author}{\bibfnamefont{R.}~\bibnamefont{Huber}},
  \bibinfo{journal}{Nano Letters} \textbf{\bibinfo{volume}{17}},
  \bibinfo{pages}{1455} (\bibinfo{year}{2017}), \bibinfo{note}{pMID: 28182430},
  \eprint{https://doi.org/10.1021/acs.nanolett.6b04422},
  \urlprefix\url{https://doi.org/10.1021/acs.nanolett.6b04422}.

\bibitem[{\citenamefont{Cunningham et~al.}(2017)\citenamefont{Cunningham,
  Hanbicki, McCreary, and Jonker}}]{Cunningham-ACSNano2017}
\bibinfo{author}{\bibfnamefont{P.~D.} \bibnamefont{Cunningham}},
  \bibinfo{author}{\bibfnamefont{A.~T.} \bibnamefont{Hanbicki}},
  \bibinfo{author}{\bibfnamefont{K.~M.} \bibnamefont{McCreary}},
  \bibnamefont{and} \bibinfo{author}{\bibfnamefont{B.~T.}
  \bibnamefont{Jonker}}, \bibinfo{journal}{ACS Nano}
  \textbf{\bibinfo{volume}{11}}, \bibinfo{pages}{12601} (\bibinfo{year}{2017}),
  \bibinfo{note}{pMID: 29227085},
  \eprint{https://doi.org/10.1021/acsnano.7b06885},
  \urlprefix\url{https://doi.org/10.1021/acsnano.7b06885}.

\bibitem[{\citenamefont{Sie et~al.}(2017)\citenamefont{Sie, Steinhoff, Gies,
  Lui, Ma, Rösner, Schönhoff, Jahnke, Wehling, Lee
  et~al.}}]{Sie-NanoLett2017}
\bibinfo{author}{\bibfnamefont{E.~J.} \bibnamefont{Sie}},
  \bibinfo{author}{\bibfnamefont{A.}~\bibnamefont{Steinhoff}},
  \bibinfo{author}{\bibfnamefont{C.}~\bibnamefont{Gies}},
  \bibinfo{author}{\bibfnamefont{C.~H.} \bibnamefont{Lui}},
  \bibinfo{author}{\bibfnamefont{Q.}~\bibnamefont{Ma}},
  \bibinfo{author}{\bibfnamefont{M.}~\bibnamefont{Rösner}},
  \bibinfo{author}{\bibfnamefont{G.}~\bibnamefont{Schönhoff}},
  \bibinfo{author}{\bibfnamefont{F.}~\bibnamefont{Jahnke}},
  \bibinfo{author}{\bibfnamefont{T.~O.} \bibnamefont{Wehling}},
  \bibinfo{author}{\bibfnamefont{Y.-H.} \bibnamefont{Lee}},
  \bibnamefont{et~al.}, \bibinfo{journal}{Nano Letters}
  \textbf{\bibinfo{volume}{17}}, \bibinfo{pages}{4210} (\bibinfo{year}{2017}),
  \bibinfo{note}{pMID: 28621953},
  \eprint{https://doi.org/10.1021/acs.nanolett.7b01034},
  \urlprefix\url{https://doi.org/10.1021/acs.nanolett.7b01034}.

\bibitem[{\citenamefont{Ruppert et~al.}(2017)\citenamefont{Ruppert, Chernikov,
  Hill, Rigosi, and Heinz}}]{Ruppert-NanoLett2017}
\bibinfo{author}{\bibfnamefont{C.}~\bibnamefont{Ruppert}},
  \bibinfo{author}{\bibfnamefont{A.}~\bibnamefont{Chernikov}},
  \bibinfo{author}{\bibfnamefont{H.~M.} \bibnamefont{Hill}},
  \bibinfo{author}{\bibfnamefont{A.~F.} \bibnamefont{Rigosi}},
  \bibnamefont{and} \bibinfo{author}{\bibfnamefont{T.~F.} \bibnamefont{Heinz}},
  \bibinfo{journal}{Nano Letters} \textbf{\bibinfo{volume}{17}},
  \bibinfo{pages}{644} (\bibinfo{year}{2017}), \bibinfo{note}{pMID: 28059520},
  \eprint{https://doi.org/10.1021/acs.nanolett.6b03513},
  \urlprefix\url{https://doi.org/10.1021/acs.nanolett.6b03513}.

\bibitem[{\citenamefont{Jiang et~al.}(2018)\citenamefont{Jiang, Chen, Zheng,
  Xu, and Tang}}]{Jiang-Opt2018}
\bibinfo{author}{\bibfnamefont{T.}~\bibnamefont{Jiang}},
  \bibinfo{author}{\bibfnamefont{R.}~\bibnamefont{Chen}},
  \bibinfo{author}{\bibfnamefont{X.}~\bibnamefont{Zheng}},
  \bibinfo{author}{\bibfnamefont{Z.}~\bibnamefont{Xu}}, \bibnamefont{and}
  \bibinfo{author}{\bibfnamefont{Y.}~\bibnamefont{Tang}},
  \bibinfo{journal}{Opt. Express} \textbf{\bibinfo{volume}{26}},
  \bibinfo{pages}{859} (\bibinfo{year}{2018}),
  \urlprefix\url{http://www.opticsexpress.org/abstract.cfm?URI=oe-26-2-859}.

\bibitem[{\citenamefont{Bataller et~al.}(2019)\citenamefont{Bataller, Younts,
  Rustagi, Yu, Ardekani, Kemper, Cao, and Gundogdu}}]{Bataller-NanoLett2019}
\bibinfo{author}{\bibfnamefont{A.~W.} \bibnamefont{Bataller}},
  \bibinfo{author}{\bibfnamefont{R.~A.} \bibnamefont{Younts}},
  \bibinfo{author}{\bibfnamefont{A.}~\bibnamefont{Rustagi}},
  \bibinfo{author}{\bibfnamefont{Y.}~\bibnamefont{Yu}},
  \bibinfo{author}{\bibfnamefont{H.}~\bibnamefont{Ardekani}},
  \bibinfo{author}{\bibfnamefont{A.}~\bibnamefont{Kemper}},
  \bibinfo{author}{\bibfnamefont{L.}~\bibnamefont{Cao}}, \bibnamefont{and}
  \bibinfo{author}{\bibfnamefont{K.}~\bibnamefont{Gundogdu}},
  \bibinfo{journal}{Nano Letters} \textbf{\bibinfo{volume}{19}},
  \bibinfo{pages}{1104} (\bibinfo{year}{2019}),
  \eprint{https://doi.org/10.1021/acs.nanolett.8b04408},
  \urlprefix\url{https://doi.org/10.1021/acs.nanolett.8b04408}.

\bibitem[{\citenamefont{Kaindl et~al.}(2003)\citenamefont{Kaindl, Carnahan,
  H{\"a}gele, L{\"o}venich, and Chemla}}]{Kaindl-Nature2003}
\bibinfo{author}{\bibfnamefont{R.~A.} \bibnamefont{Kaindl}},
  \bibinfo{author}{\bibfnamefont{M.~A.} \bibnamefont{Carnahan}},
  \bibinfo{author}{\bibfnamefont{D.}~\bibnamefont{H{\"a}gele}},
  \bibinfo{author}{\bibfnamefont{R.}~\bibnamefont{L{\"o}venich}},
  \bibnamefont{and} \bibinfo{author}{\bibfnamefont{D.~S.}
  \bibnamefont{Chemla}}, \bibinfo{journal}{Nature}
  \textbf{\bibinfo{volume}{423}}, \bibinfo{pages}{734 EP }
  (\bibinfo{year}{2003}),
  \urlprefix\url{http://dx.doi.org/10.1038/nature01676}.

\bibitem[{\citenamefont{Nozi{\`e}res and Schmitt-Rink}(1985)}]{Nozieres1985}
\bibinfo{author}{\bibfnamefont{P.}~\bibnamefont{Nozi{\`e}res}}
  \bibnamefont{and}
  \bibinfo{author}{\bibfnamefont{S.}~\bibnamefont{Schmitt-Rink}},
  \bibinfo{journal}{Journal of Low Temperature Physics}
  \textbf{\bibinfo{volume}{59}}, \bibinfo{pages}{195} (\bibinfo{year}{1985}),
  ISSN \bibinfo{issn}{1573-7357},
  \urlprefix\url{http://dx.doi.org/10.1007/BF00683774}.

\bibitem[{\citenamefont{Capone et~al.}(2002)\citenamefont{Capone, Castellani,
  and Grilli}}]{Massimo02}
\bibinfo{author}{\bibfnamefont{M.}~\bibnamefont{Capone}},
  \bibinfo{author}{\bibfnamefont{C.}~\bibnamefont{Castellani}},
  \bibnamefont{and} \bibinfo{author}{\bibfnamefont{M.}~\bibnamefont{Grilli}},
  \bibinfo{journal}{Phys. Rev. Lett.} \textbf{\bibinfo{volume}{88}},
  \bibinfo{pages}{126403} (\bibinfo{year}{2002}), \bibinfo{note}{actually study
  the attractive case at fixed density, which, as mentioned in the text, is
  equivalent to the repulsive one at fixed magnetisation},
  \urlprefix\url{https://link.aps.org/doi/10.1103/PhysRevLett.88.126403}.

\bibitem[{\citenamefont{Moreo and Scalapino}(2007)}]{Scalapino2007}
\bibinfo{author}{\bibfnamefont{A.}~\bibnamefont{Moreo}} \bibnamefont{and}
  \bibinfo{author}{\bibfnamefont{D.~J.} \bibnamefont{Scalapino}},
  \bibinfo{journal}{Phys. Rev. Lett.} \textbf{\bibinfo{volume}{98}},
  \bibinfo{pages}{216402} (\bibinfo{year}{2007}),
  \urlprefix\url{https://link.aps.org/doi/10.1103/PhysRevLett.98.216402}.

\bibitem[{\citenamefont{Lieb}(1989)}]{Lieb89}
\bibinfo{author}{\bibfnamefont{E.~H.} \bibnamefont{Lieb}},
  \bibinfo{journal}{Phys. Rev. Lett.} \textbf{\bibinfo{volume}{62}},
  \bibinfo{pages}{1201} (\bibinfo{year}{1989}),
  \urlprefix\url{https://link.aps.org/doi/10.1103/PhysRevLett.62.1201}.

\bibitem[{\citenamefont{\ifmmode~\check{Z}\else \v{Z}\fi{}itko
  et~al.}(2015)\citenamefont{\ifmmode~\check{Z}\else \v{Z}\fi{}itko, Osolin,
  and Jegli\ifmmode~\check{c}\else \v{c}\fi{}}}]{Rok_Osolin_2015}
\bibinfo{author}{\bibfnamefont{R.}~\bibnamefont{\ifmmode~\check{Z}\else
  \v{Z}\fi{}itko}}, \bibinfo{author}{\bibfnamefont{i.~c.~v.}
  \bibnamefont{Osolin}}, \bibnamefont{and}
  \bibinfo{author}{\bibfnamefont{P.}~\bibnamefont{Jegli\ifmmode~\check{c}\else
  \v{c}\fi{}}}, \bibinfo{journal}{Phys. Rev. B} \textbf{\bibinfo{volume}{91}},
  \bibinfo{pages}{155111} (\bibinfo{year}{2015}),
  \urlprefix\url{https://link.aps.org/doi/10.1103/PhysRevB.91.155111}.

\bibitem[{\citenamefont{Georges et~al.}(1996)\citenamefont{Georges, Kotliar,
  Krauth, and Rozenberg}}]{ReviewDMFT}
\bibinfo{author}{\bibfnamefont{A.}~\bibnamefont{Georges}},
  \bibinfo{author}{\bibfnamefont{G.}~\bibnamefont{Kotliar}},
  \bibinfo{author}{\bibfnamefont{W.}~\bibnamefont{Krauth}}, \bibnamefont{and}
  \bibinfo{author}{\bibfnamefont{M.~J.} \bibnamefont{Rozenberg}},
  \bibinfo{journal}{Rev. Mod. Phys.} \textbf{\bibinfo{volume}{68}},
  \bibinfo{pages}{13} (\bibinfo{year}{1996}).

\bibitem[{\citenamefont{Laloux et~al.}(1994)\citenamefont{Laloux, Georges, and
  Krauth}}]{GeorgesPRB1994}
\bibinfo{author}{\bibfnamefont{L.}~\bibnamefont{Laloux}},
  \bibinfo{author}{\bibfnamefont{A.}~\bibnamefont{Georges}}, \bibnamefont{and}
  \bibinfo{author}{\bibfnamefont{W.}~\bibnamefont{Krauth}},
  \bibinfo{journal}{Phys. Rev. B} \textbf{\bibinfo{volume}{50}},
  \bibinfo{pages}{3092} (\bibinfo{year}{1994}),
  \urlprefix\url{https://link.aps.org/doi/10.1103/PhysRevB.50.3092}.

\bibitem[{\citenamefont{Keller et~al.}(2001)\citenamefont{Keller, Metzner, and
  Schollw\"ock}}]{Metzner01}
\bibinfo{author}{\bibfnamefont{M.}~\bibnamefont{Keller}},
  \bibinfo{author}{\bibfnamefont{W.}~\bibnamefont{Metzner}}, \bibnamefont{and}
  \bibinfo{author}{\bibfnamefont{U.}~\bibnamefont{Schollw\"ock}},
  \bibinfo{journal}{Phys. Rev. Lett.} \textbf{\bibinfo{volume}{86}},
  \bibinfo{pages}{4612} (\bibinfo{year}{2001}),
  \urlprefix\url{https://link.aps.org/doi/10.1103/PhysRevLett.86.4612}.

\bibitem[{\citenamefont{Bauer and Hewson}(2007)}]{HewsonPRB2007}
\bibinfo{author}{\bibfnamefont{J.}~\bibnamefont{Bauer}} \bibnamefont{and}
  \bibinfo{author}{\bibfnamefont{A.~C.} \bibnamefont{Hewson}},
  \bibinfo{journal}{Phys. Rev. B} \textbf{\bibinfo{volume}{76}},
  \bibinfo{pages}{035118} (\bibinfo{year}{2007}),
  \urlprefix\url{https://link.aps.org/doi/10.1103/PhysRevB.76.035118}.

\bibitem[{\citenamefont{Zhu et~al.}(2017)\citenamefont{Zhu, Sheng, and
  Zhu}}]{Zhu_2017}
\bibinfo{author}{\bibfnamefont{W.}~\bibnamefont{Zhu}},
  \bibinfo{author}{\bibfnamefont{D.~N.} \bibnamefont{Sheng}}, \bibnamefont{and}
  \bibinfo{author}{\bibfnamefont{J.-X.} \bibnamefont{Zhu}},
  \bibinfo{journal}{Phys. Rev. B} \textbf{\bibinfo{volume}{96}},
  \bibinfo{pages}{085118} (\bibinfo{year}{2017}),
  \urlprefix\url{https://link.aps.org/doi/10.1103/PhysRevB.96.085118}.

\bibitem[{\citenamefont{Kotliar}(1999)}]{Kotliar1999}
\bibinfo{author}{\bibfnamefont{G.}~\bibnamefont{Kotliar}},
  \bibinfo{journal}{The European Physical Journal B - Condensed Matter and
  Complex Systems} \textbf{\bibinfo{volume}{11}}, \bibinfo{pages}{27}
  (\bibinfo{year}{1999}), ISSN \bibinfo{issn}{1434-6036},
  \urlprefix\url{http://dx.doi.org/10.1007/s100510050914}.

\bibitem[{\citenamefont{Lanat\`a et~al.}(2017)\citenamefont{Lanat\`a, Lee, Yao,
  and Dobrosavljevi\ifmmode~\acute{c}\else \'{c}\fi{}}}]{Lanata-ghost-2017}
\bibinfo{author}{\bibfnamefont{N.}~\bibnamefont{Lanat\`a}},
  \bibinfo{author}{\bibfnamefont{T.-H.} \bibnamefont{Lee}},
  \bibinfo{author}{\bibfnamefont{Y.-X.} \bibnamefont{Yao}}, \bibnamefont{and}
  \bibinfo{author}{\bibfnamefont{V.}~\bibnamefont{Dobrosavljevi\ifmmode~\acute{c}\else
  \'{c}\fi{}}}, \bibinfo{journal}{Phys. Rev. B} \textbf{\bibinfo{volume}{96}},
  \bibinfo{pages}{195126} (\bibinfo{year}{2017}),
  \urlprefix\url{https://link.aps.org/doi/10.1103/PhysRevB.96.195126}.

\bibitem[{\citenamefont{Gutzwiller}(1963)}]{Gutzwiller-1}
\bibinfo{author}{\bibfnamefont{M.~C.} \bibnamefont{Gutzwiller}},
  \bibinfo{journal}{Phys. Rev. Lett.} \textbf{\bibinfo{volume}{10}},
  \bibinfo{pages}{159} (\bibinfo{year}{1963}),
  \urlprefix\url{http://link.aps.org/doi/10.1103/PhysRevLett.10.159}.

\bibitem[{\citenamefont{Gutzwiller}(1965)}]{Gutzwiller-2}
\bibinfo{author}{\bibfnamefont{M.~C.} \bibnamefont{Gutzwiller}},
  \bibinfo{journal}{Phys. Rev.} \textbf{\bibinfo{volume}{137}},
  \bibinfo{pages}{A1726} (\bibinfo{year}{1965}).

\bibitem[{\citenamefont{Vitiello et~al.}(1988)\citenamefont{Vitiello, Runge,
  and Kalos}}]{Shadow-WF}
\bibinfo{author}{\bibfnamefont{S.}~\bibnamefont{Vitiello}},
  \bibinfo{author}{\bibfnamefont{K.}~\bibnamefont{Runge}}, \bibnamefont{and}
  \bibinfo{author}{\bibfnamefont{M.~H.} \bibnamefont{Kalos}},
  \bibinfo{journal}{Phys. Rev. Lett.} \textbf{\bibinfo{volume}{60}},
  \bibinfo{pages}{1970} (\bibinfo{year}{1988}),
  \urlprefix\url{https://link.aps.org/doi/10.1103/PhysRevLett.60.1970}.

\bibitem[{\citenamefont{Rommer and \"Ostlund}(1997)}]{MPS-1}
\bibinfo{author}{\bibfnamefont{S.}~\bibnamefont{Rommer}} \bibnamefont{and}
  \bibinfo{author}{\bibfnamefont{S.}~\bibnamefont{\"Ostlund}},
  \bibinfo{journal}{Phys. Rev. B} \textbf{\bibinfo{volume}{55}},
  \bibinfo{pages}{2164} (\bibinfo{year}{1997}),
  \urlprefix\url{https://link.aps.org/doi/10.1103/PhysRevB.55.2164}.

\bibitem[{\citenamefont{Or\'us}(2014)}]{MPS-2}
\bibinfo{author}{\bibfnamefont{R.}~\bibnamefont{Or\'us}},
  \bibinfo{journal}{Annals of Physics} \textbf{\bibinfo{volume}{349}},
  \bibinfo{pages}{117 } (\bibinfo{year}{2014}), ISSN \bibinfo{issn}{0003-4916},
  \urlprefix\url{http://www.sciencedirect.com/science/article/pii/S0003491614001596}.

\bibitem[{\citenamefont{Carleo and Troyer}(2017)}]{Carleo-Science2017}
\bibinfo{author}{\bibfnamefont{G.}~\bibnamefont{Carleo}} \bibnamefont{and}
  \bibinfo{author}{\bibfnamefont{M.}~\bibnamefont{Troyer}},
  \bibinfo{journal}{Science} \textbf{\bibinfo{volume}{355}},
  \bibinfo{pages}{602} (\bibinfo{year}{2017}), ISSN \bibinfo{issn}{0036-8075},
  \eprint{http://science.sciencemag.org/content/355/6325/602.full.pdf},
  \urlprefix\url{http://science.sciencemag.org/content/355/6325/602}.

\bibitem[{\citenamefont{Lanat\`a et~al.}(2015)\citenamefont{Lanat\`a, Yao,
  Wang, Ho, and Kotliar}}]{NicolaPRX}
\bibinfo{author}{\bibfnamefont{N.}~\bibnamefont{Lanat\`a}},
  \bibinfo{author}{\bibfnamefont{Y.}~\bibnamefont{Yao}},
  \bibinfo{author}{\bibfnamefont{C.-Z.} \bibnamefont{Wang}},
  \bibinfo{author}{\bibfnamefont{K.-M.} \bibnamefont{Ho}}, \bibnamefont{and}
  \bibinfo{author}{\bibfnamefont{G.}~\bibnamefont{Kotliar}},
  \bibinfo{journal}{Phys. Rev. X} \textbf{\bibinfo{volume}{5}},
  \bibinfo{pages}{011008} (\bibinfo{year}{2015}),
  \urlprefix\url{http://link.aps.org/doi/10.1103/PhysRevX.5.011008}.

\bibitem[{\citenamefont{B\"unemann et~al.}(1998)\citenamefont{B\"unemann,
  Weber, and Gebhard}}]{BW&G1998}
\bibinfo{author}{\bibfnamefont{J.}~\bibnamefont{B\"unemann}},
  \bibinfo{author}{\bibfnamefont{W.}~\bibnamefont{Weber}}, \bibnamefont{and}
  \bibinfo{author}{\bibfnamefont{F.}~\bibnamefont{Gebhard}},
  \bibinfo{journal}{Phys. Rev. B} \textbf{\bibinfo{volume}{57}},
  \bibinfo{pages}{6896} (\bibinfo{year}{1998}),
  \urlprefix\url{http://link.aps.org/doi/10.1103/PhysRevB.57.6896}.

\bibitem[{\citenamefont{Fabrizio}(2017)}]{Michele17}
\bibinfo{author}{\bibfnamefont{M.}~\bibnamefont{Fabrizio}},
  \bibinfo{journal}{Phys. Rev. B} \textbf{\bibinfo{volume}{95}},
  \bibinfo{pages}{075156} (\bibinfo{year}{2017}),
  \urlprefix\url{https://link.aps.org/doi/10.1103/PhysRevB.95.075156}.

\bibitem[{\citenamefont{Garc\'{\i}a et~al.}(2004)\citenamefont{Garc\'{\i}a,
  Hallberg, and Rozenberg}}]{Hallberg2004}
\bibinfo{author}{\bibfnamefont{D.~J.} \bibnamefont{Garc\'{\i}a}},
  \bibinfo{author}{\bibfnamefont{K.}~\bibnamefont{Hallberg}}, \bibnamefont{and}
  \bibinfo{author}{\bibfnamefont{M.~J.} \bibnamefont{Rozenberg}},
  \bibinfo{journal}{Phys. Rev. Lett.} \textbf{\bibinfo{volume}{93}},
  \bibinfo{pages}{246403} (\bibinfo{year}{2004}),
  \urlprefix\url{https://link.aps.org/doi/10.1103/PhysRevLett.93.246403}.

\bibitem[{\citenamefont{Werner and Millis}(2007)}]{Werner2007}
\bibinfo{author}{\bibfnamefont{P.}~\bibnamefont{Werner}} \bibnamefont{and}
  \bibinfo{author}{\bibfnamefont{A.~J.} \bibnamefont{Millis}},
  \bibinfo{journal}{Phys. Rev. B} \textbf{\bibinfo{volume}{75}},
  \bibinfo{pages}{085108} (\bibinfo{year}{2007}),
  \urlprefix\url{https://link.aps.org/doi/10.1103/PhysRevB.75.085108}.

\bibitem[{\citenamefont{Bulla}(1999)}]{BullaPRL1999}
\bibinfo{author}{\bibfnamefont{R.}~\bibnamefont{Bulla}},
  \bibinfo{journal}{Phys. Rev. Lett.} \textbf{\bibinfo{volume}{83}},
  \bibinfo{pages}{136} (\bibinfo{year}{1999}),
  \urlprefix\url{https://link.aps.org/doi/10.1103/PhysRevLett.83.136}.

\bibitem[{\citenamefont{Tong et~al.}(2001)\citenamefont{Tong, Shen, and
  Pu}}]{PuPRB2001}
\bibinfo{author}{\bibfnamefont{N.-H.} \bibnamefont{Tong}},
  \bibinfo{author}{\bibfnamefont{S.-Q.} \bibnamefont{Shen}}, \bibnamefont{and}
  \bibinfo{author}{\bibfnamefont{F.-C.} \bibnamefont{Pu}},
  \bibinfo{journal}{Phys. Rev. B} \textbf{\bibinfo{volume}{64}},
  \bibinfo{pages}{235109} (\bibinfo{year}{2001}),
  \urlprefix\url{https://link.aps.org/doi/10.1103/PhysRevB.64.235109}.

\bibitem[{\citenamefont{Grivickas et~al.}(2003)\citenamefont{Grivickas,
  Grivickas, and Linnros}}]{Grivickas-PRL2003}
\bibinfo{author}{\bibfnamefont{P.}~\bibnamefont{Grivickas}},
  \bibinfo{author}{\bibfnamefont{V.}~\bibnamefont{Grivickas}},
  \bibnamefont{and} \bibinfo{author}{\bibfnamefont{J.}~\bibnamefont{Linnros}},
  \bibinfo{journal}{Phys. Rev. Lett.} \textbf{\bibinfo{volume}{91}},
  \bibinfo{pages}{246401} (\bibinfo{year}{2003}),
  \urlprefix\url{https://link.aps.org/doi/10.1103/PhysRevLett.91.246401}.

\bibitem[{\citenamefont{Sekiguchi et~al.}(2017)\citenamefont{Sekiguchi,
  Mochizuki, Kim, Akiyama, Pfeiffer, West, and Shimano}}]{Sekiguchi-PRL2017}
\bibinfo{author}{\bibfnamefont{F.}~\bibnamefont{Sekiguchi}},
  \bibinfo{author}{\bibfnamefont{T.}~\bibnamefont{Mochizuki}},
  \bibinfo{author}{\bibfnamefont{C.}~\bibnamefont{Kim}},
  \bibinfo{author}{\bibfnamefont{H.}~\bibnamefont{Akiyama}},
  \bibinfo{author}{\bibfnamefont{L.~N.} \bibnamefont{Pfeiffer}},
  \bibinfo{author}{\bibfnamefont{K.~W.} \bibnamefont{West}}, \bibnamefont{and}
  \bibinfo{author}{\bibfnamefont{R.}~\bibnamefont{Shimano}},
  \bibinfo{journal}{Phys. Rev. Lett.} \textbf{\bibinfo{volume}{118}},
  \bibinfo{pages}{067401} (\bibinfo{year}{2017}),
  \urlprefix\url{https://link.aps.org/doi/10.1103/PhysRevLett.118.067401}.

\bibitem[{\citenamefont{Mahan}(1967)}]{Mahan_67}
\bibinfo{author}{\bibfnamefont{G.~D.} \bibnamefont{Mahan}},
  \bibinfo{journal}{Phys. Rev.} \textbf{\bibinfo{volume}{153}},
  \bibinfo{pages}{882} (\bibinfo{year}{1967}),
  \urlprefix\url{https://link.aps.org/doi/10.1103/PhysRev.153.882}.

\bibitem[{\citenamefont{Smith and Wolfe}(1995)}]{Wolfe-PRB1995}
\bibinfo{author}{\bibfnamefont{L.~M.} \bibnamefont{Smith}} \bibnamefont{and}
  \bibinfo{author}{\bibfnamefont{J.~P.} \bibnamefont{Wolfe}},
  \bibinfo{journal}{Phys. Rev. B} \textbf{\bibinfo{volume}{51}},
  \bibinfo{pages}{7521} (\bibinfo{year}{1995}),
  \urlprefix\url{https://link.aps.org/doi/10.1103/PhysRevB.51.7521}.

\end{thebibliography}
\end{document}